\documentclass[twocolumn,english,preprintnumbers,amsmath,amssymb,pre,longbibliography]{revtex4-1}

\usepackage{graphicx}
\usepackage{color}

\begin{document}

\title{Elastic properties of cubic silicon carbide with Si vacancies}

\author{Carlos P. Herrero$^1$\footnote{Electronic mail: ch@icmm.csic.es},
    Eduardo R. Hern\'andez$^1$, and Gabriela Herrero-Saboya$^2$}

\affiliation{$^1$Instituto de Ciencia de Materiales de Madrid,
           Consejo Superior de Investigaciones Cient\'ificas (CSIC),
           Campus de Cantoblanco, 28049 Madrid, Spain  \\
         $^2$CNR-IOM Democritos National Simulation Center,
           Istituto Officina dei Materiali, c/o SISSA,
           via Bonomea 265, IT-34136 Trieste, Italy}	 

\date{\today}

\begin{abstract}
We investigate how silicon vacancies modify the elastic response and
mechanical stability of cubic $3C$-SiC. Our approach employs 
path-integral molecular dynamics simulations, including the
classical-nuclei limit,	based on an efficient 
tight-binding Hamiltonian, whose accuracy is validated against 
density-functional-theory calculations. This framework enables 
a quantitative assessment of nuclear quantum effects arising from 
zero-point motion. Across a broad range of temperatures and hydrostatic 
pressures, spanning both tensile and compressive regimes, silicon vacancies 
are found to substantially renormalize the elastic constants 
$C_{11}$, $C_{12}$, and $C_{44}$, as well as the bulk modulus, 
relative to the defect-free crystal. Inclusion of nuclear quantum motion 
produces an additional softening of these elastic properties, particularly 
at low temperatures, demonstrating that quantum fluctuations make 
a measurable contribution to the mechanical response of defective SiC.
Vacancies also affect the mechanical stability domain of $3C$-SiC, lowering 
the maximum sustainable tensile pressure by approximately 4~GPa for a defect
concentration of 0.016 per lattice site. These results reveal an interplay 
between point defects and quantum lattice fluctuations in determining 
the elastic behavior of SiC, providing microscopic insight relevant for 
both extreme-environment structural applications and defect-based 
quantum technologies.
\end{abstract}

\maketitle

\section{Introduction}

Bulk silicon carbide is a semiconducting material with exceptional 
properties, such as high thermal conductivity, strength, and refractive 
index, as well as low thermal expansion, which makes it attractive
for high-temperature and 
high-power electronics \cite{sc-sh17,sc-pa22}.
The elastic properties of SiC have been studied
over the years by theoretical
\cite{sc-va15,sc-ra21,sc-pe22,sc-sh00,sc-he23}
and experimental approaches \cite{sc-zh13,sc-le82,sc-la91},
because of their importance in both basic research and
technological applications.
In particular, the cubic phase $3C$-SiC, stable under ambient
conditions, is one of the most studied polymorphs of silicon carbide,
and its mechanical properties have been extensively analyzed
\cite{sc-le82,sc-sh00,sc-zh13,sc-ni17,sc-da18,sc-ra21,sc-pe22,sc-he23}.

Silicon carbide has emerged as a promising platform for quantum 
technologies because it hosts optically addressable spin defects, 
including the silicon vacancy ($V_{\rm Si}$) in various polymorphs 
\cite{sc-ko11,sc-sh20,sc-ud20,sc-wi19}. 
These defects arise from missing Si atoms and possess
localized electron spins that can be optically initialized, controlled,
and read out, with long coherence times under suitable protocols
\cite{sc-go15,sc-iv17,sc-le21,sc-fa24,sc-zh23}.
$V_{\rm Si}$ centers emit near-infrared photons applicable for 
fiber-based communication, and enable finely-tuned nanoscale
sensing of fields, temperature, and strain.

The $V_{\rm Si}$ defect in cubic SiC has been investigated 
by means of several theoretical methods, including
{\em ab initio} density-functional-theory (DFT) calculations
\cite{sc-de99,sc-bo03,sc-ru03b,sc-sc21,sc-qi23} 
and finite-temperature atomistic simulations, especially molecular 
dynamics (MD) \cite{sc-sa04,sc-ma00,sc-li19,sc-le21,sc-ra24}.
These studies have provided insight into the structural, electronic, 
optical, and dynamical properties of silicon vacancies in this 
material  \cite{sc-zy99,sc-sa04,sc-le21,sc-fa22,sc-zh24}.
In contrast, mechanical aspects of vacancy-containing SiC, 
particularly its elastic response, have received less attention 
\cite{sc-li19,sc-fa22,sc-qi23,sc-ra24}.
MD simulations have addressed defect energetics \cite{sc-sa04},
stability \cite{sc-le21}, and mechanical properties of the defective
solid \cite{sc-li19,sc-ra24}, as well as
irradiation-induced defect creation \cite{sc-ma00,sc-li19}.
In particular, Li and Xiao \cite{sc-li19} studied the influence
of point defects on the tensile strength, {\em i.e.}, the maximum 
stress the material can endure when being stretched before it 
fractures.  In this context, a precise determination of the 
mechanical instability threshold is limited by the presence of 
metastable configurations during the material deformation.

In this paper, we use MD simulations to study the influence 
of neutral silicon vacancies on the elastic properties 
of $3C$-SiC over the temperature range $T = 50$ to 
1200~K, and hydrostatic pressures from 
$P = -44$~GPa (tension) to 60~GPa (compression). 
We describe the interatomic interactions through an efficient
tight-binding (TB) Hamiltonian, whose accuracy is validated by
DFT calculations at $T = 0$ \cite{po95,go97,co05}.
The inclusion of negative pressures allows us to approach the 
mechanical stability limit of the material, and to analyze the 
shift of the spinodal point $P_s$ in the presence of vacancies.
Within this framework, the combination of trustworthy electronic 
structure calculations with finite-temperature MD simulations 
provides an efficient groundwork to analyze the interplay between
temperature, stress, and structure in vacancy-containing SiC
\cite{sc-he25}.
The outcome of our simulations gives helpful insight on the
elastic properties of $3C$-SiC. 
To assess the effect of nuclear quantum motion (for both C and Si),
we have also carried out path-integral molecular dynamics (PIMD)
simulations based on the same TB-derived interactions 
\cite{ce95,he14,ca17,sc-br25}.
According to our results, such quantum corrections cause in some
cases appreciable changes in the elastic constants of the 
vacancy-containing solid at relatively low temperatures. 

Our main goal in this paper is threefold. 
First, we focus on the influence of temperature and pressure
on the elastic properties of defective SiC, and in particular
on the effect of Si vacancies upon its mechanical stability
from a firm thermodynamic basis.
Second, we concentrate on the response of physical features,
{\em e.g.} bulk modulus, to the defect concentration.
Third, we analyze the effect of nuclear quantum motion on
elastic and structural properties, with especial emphasis 
on the induced softening of the material, usually neglected 
in this kind of calculations.

The paper is organized as follows: in Sec.~II we present the 
computational methods employed, including molecular dynamics 
simulations, tight-binding method, and DFT procedure.
In Sec.~III, we outline the results and discussion for the
energy (III.A), atomic mean-square displacements (III.B), 
volume (III.C), elastic constants (III.D), and bulk modulus 
(III.E).  The main findings are summarized in Sec.~IV.

\section{Method of calculation}

We investigate structural and mechanical properties of $3C$ silicon 
carbide containing Si vacancies by means of two types of molecular dynamics 
simulations, which allow us to determine the equilibrium states of 
this system under different temperature and pressure conditions.
On the one hand, we perform classical MD simulations, in which the atomic 
motion is governed by Newton's equations of motion, numerically integrated 
over time. On the other hand, we employ PIMD simulations in order to 
explicitly account for the quantum nature of the atomic nuclei.

From a computational perspective, the main difference between these two 
approaches lies in the representation of the atomic nuclei.
In PIMD simulations, each nucleus is modeled as an ensemble of $N_{\rm Tr}$ 
replicas (Trotter number), which behave as classical particles (beads) 
arranged in a ring-polymer configuration \cite{fe72,gi88,ce95,he14}.
This mapping gives rise to a pseudoclassical system that provides accurate 
equilibrium quantum-mechanical properties.
By comparing results obtained from classical MD and PIMD simulations, one 
can quantitatively assess the magnitude of nuclear quantum effects in various 
physical observables.
Within this framework, the classical limit is recovered by setting 
$N_{\rm Tr} = 1$, for which each ring polymer collapses into a single 
particle, and quantum delocalization effects are absent.

An important issue in our finite-temperature simulations concerns the 
choice of interatomic interactions, which must be described as 
realistically as possible.
In principle, this could be achieved by employing {\em ab initio} 
density-functional theory or Hartree-Fock self-consistent potentials.
However, such approaches would severely limit the accessible simulation 
times and/or the size of the simulation cell within reasonable computational 
resources.  For this reason, we derive the interatomic interactions from 
an efficient non-orthogonal TB Hamiltonian, parameterized 
on the basis of DFT calculations \cite{po95}.
This class of TB methods has been shown to provide reliable results for 
a wide range of properties in both condensed-matter and molecular 
systems \cite{go97,co05}.

The TB formalism employed in our MD simulations is adapted from the 
TROCADERO package \cite{si-ru03}.
The parameterization for systems containing C and Si atoms was developed 
in Refs.~\cite{po95,gu96}, and has been successfully applied to the study 
of bulk silicon carbide \cite{ra08,sc-me96,sc-be05}, isotopic and nuclear 
quantum effects in $3C$-SiC \cite{he09c,sc-he25,sc-he24}, and surface 
reconstructions of this material \cite{gu96}.
More recently, it has also been used to investigate various physical 
properties of newly synthesized SiC monolayers \cite{he22,sc-po23}.
A comprehensive discussion of the capability of this type of TB approach 
to describe diverse properties of molecular and condensed-matter systems 
can be found in the work of Goringe {\it et al.} \cite{go97}.
We emphasize that our ``classical'' MD simulations are based on the 
TB-derived interatomic interaction rather than on an empirical effective 
potential. Thus, the term ``classical'' refers exclusively to the
treatment of the nuclear dynamics, as noted above.

Our simulations were performed in the isothermal-isobaric ($NPT$) ensemble, 
using algorithms based on well-established methods reported in the 
literature \cite{tu92,ma99}.
In the PIMD simulations, staging coordinates were employed to represent 
the bead positions within the ring polymers.
Each staging coordinate was coupled to a chain of four Nos\'e-Hoover 
thermostats to maintain a constant temperature.
In addition, the barostat was connected to a chain of four thermostats, 
allowing the volume fluctuations required to sample the target 
pressure \cite{tu10,he14}.
The equations of motion were integrated numerically using the reversible 
reference system propagator algorithm (RESPA), which enables the use of 
multiple time steps for fast and slow dynamical variables \cite{ma96}.
For the fast degrees of freedom, such as thermostat variables and 
bead-bead interactions, we employed a time step of $\delta t = 0.25$~fs.
For the slower dynamics associated with the interatomic forces, a larger 
time step of $\Delta t = 1$~fs was used.
Further details of the simulation methodology can be found elsewhere 
\cite{tu10,he16}.

Our simulations, including both classical MD and PIMD,
were primarily performed using $2 \times 2 \times 2$ and
$3 \times 3 \times 3$ supercells of the face-centered cubic unit cell of
$3C$-SiC, containing $N_0 = 64$ and 216 atoms, respectively, and subject
to periodic boundary conditions.
To assess the convergence of the $T = 0$ results with respect to system
size, including the minimum energy $E_0$ and the formation energy $E_f$,
some calculations were performed using larger supercells with up
to $N_0 = 1000$ atoms.
The configurational space was sampled over a temperature range from 50
to 1200~K and under hydrostatic pressures $P$ between $-44$ and 60~GPa.
Within the framework of elasticity theory, one has
$\sigma_{xx} = \sigma_{yy} = \sigma_{zz} = -P$,
where $\sigma_{ij}$ denotes the components of the stress tensor.
Accordingly, negative and positive values of $P$ correspond to tensile
and compressive pressures, respectively.

In both classical MD and PIMD simulations, $2 \times 10^5$ time steps were
used for system equilibration. Subsequently, ensemble averages were
accumulated over $8 \times 10^6$ time steps for $N_0 = 64$,
and over $6 \times 10^6$ steps for $N_0 = 216$.
In the PIMD simulations, the Trotter number $N_{\rm Tr}$ was chosen to be
temperature dependent according to the relation $N_{\rm Tr} T = 6000$~K.
This choice ensures an approximately constant level of accuracy across the
entire temperature range considered \cite{he16}.

Near the limit of mechanical stability under tensile pressure, some 
simulations were carried out in the canonical ($NVT$) ensemble.
This approach allows access to tensile pressures closer to the spinodal 
point, where $NPT$ simulations become unstable due to the emergence of 
large volume fluctuations.

In our TB calculations, the electronic degrees of freedom in reciprocal 
space were sampled at the $\Gamma$ point (${\bf k} = 0$) only.
Tests with larger ${\bf k}$-point sets showed a small shift in the 
total energy, without significantly affecting the energy differences 
relevant to our study.  The minimum energy $E_0$ (classical limit at 
$T = 0$) exhibits a minor shift, which decreases with increasing simulation 
cell size.  A similar trend is observed for the mean energy per atom at 
finite temperatures across different cell sizes \cite{he22}.

To assess the accuracy of the tight-binding method used to describe
the elastic properties of defective $3C$-SiC,
we carried out first-principles DFT calculations at $T = 0$.
These calculations were performed to analyze aspects that
are not directly accessible from the existing literature.
We used the \textsc{Quantum ESPRESSO} package \cite{sc-gi09,sc-gi17},
with the Perdew-Burke-Ernzerhof exchange-correlation functional 
optimized for solids (PBEsol) \cite{sc-pe08}, and
projector-augmented-wave (PAW) pseudopotentials for both carbon and silicon 
 \cite{sc-ps23}.
The kinetic energy and charge density cutoffs for the plane-wave
basis set were fixed to 45~Ry and 400~Ry, respectively.
Calculations were performed for cubic SiC supercells containing 
$N_0 - 1 = 215$ atoms and a single silicon vacancy, under
periodic boundary conditions.  
The Brillouin zone was sampled at the $\Gamma$ point only.
The reference lattice parameter was taken as the optimal one for
the conventional cell, $a = 4.36$~\AA. 
Spin polarization was included to determine the ground-state
spin configuration of the silicon vacancy.

DFT has been widely used to investigate structural, thermodynamic, 
electronic, and mechanical properties of SiC 
\cite{sc-pa94b,sc-ka94,sc-ka94b,sc-ca20}.
In particular, {\em ab initio} studies of silicon vacancies have 
reported their formation energies, stable charge and spin states,
and relaxed atomic structures
\cite{sc-zy99,sc-sc21,sc-fa22,sc-le21,sc-ud20}, as well as 
diffusion pathways and migration barriers in the solid 
\cite{sc-ru03b,sc-bo03,sc-ru04,sc-de18b}.
These results provide a useful benchmark to asses how accurately
approximate approaches, such as the tight-binding method
employed in the present work, capture defect-related properties 
of $3C$-SiC.

\section{Results and discussion}

\subsection{Energetics}

In this section, we investigate the internal energy of cubic SiC containing
$V_{\rm Si}$ defects, as obtained from classical MD and PIMD simulations 
performed in the $NPT$ ensemble over a wide range of temperatures and pressures.
Before presenting the simulation results, we examine the vacancy formation
energy $E_f$, which provides a useful characterization of these point defects.
For a supercell containing $N_0$ atoms, the formation energy of a neutral
silicon vacancy $V_{\rm Si}$ is defined as \cite{sc-de18b,sc-fa24,sc-wa25}:
\begin{equation}
  E_f(V_{\rm Si}) = E(N_0 - 1) - E(N_0) + \mu_{\rm Si} \; ,
\label{ef}
\end{equation}
where $E(N_0-1)$ denotes the total energy of a supercell containing
a single Si vacancy, and $E(N_0)$ is the total energy of the
defect-free SiC supercell.
Under Si-rich conditions, $\mu_{\rm Si}$ in Eq.~(\ref{ef}) corresponds to
the chemical potential of silicon, {\em i.e.}, the energy per atom in bulk 
Si. In the C-rich (Si-poor) limit, the silicon chemical potential is given 
by $\mu_{\rm Si} = \mu_{\rm SiC} - \mu_{\rm C}$,
where $\mu_{\rm SiC} = 2 E(N_0) / N_0$ is the chemical potential of a
Si--C atom pair, and $\mu_{\rm C}$ is taken as the chemical potential of
bulk diamond \cite{sc-de18b,sc-fa24,sc-wa25}.

We use DFT calculations to establish the reference atomic structure and
formation energies of $V_{\rm Si}$ at $T = 0$, and subsequently evaluate
the accuracy of the corresponding predictions obtained with the TB Hamiltonian.
As a benchmark for our DFT calculations, we consider the negatively charged
silicon vacancy, $V_{\rm Si}^-$, which has been identified experimentally
by electron paramagnetic resonance (EPR) spectroscopy \cite{sc-it97}.
The observed EPR signal was assigned to a silicon vacancy in a quartet
electronic state with tetrahedral ($T_d$) symmetry. Consistent with this
assignment, our calculations predict a $T_d$ geometry with spin $S = 3/2$
as the ground-state configuration. The distance between the four carbon
atoms nearest to the vacancy is found to be 3.34~\AA.

Removing one electron from $V_{\rm Si}^-$ produces a slight distortion 
toward $C_{3v}$ symmetry and stabilizes a spin-triplet state ($S = 1$).
This high-spin configuration is consistent with the $S = 3/2$ ground state
of $V_{\rm Si}^-$ and is in agreement with a recent systematic study of
point defects in $3C$-SiC \cite{sc-sc21}. In the distorted structure,
one of the four nearest-neighbor carbon atoms relaxes slightly toward the
plane defined by the other three carbon atoms. At the equilibrium lattice
parameter, the displaced carbon atom is separated by 3.36~\AA\ from each of
the three coplanar neighbors, whereas the distances between the coplanar
carbon atoms are 3.40~\AA. The trigonal distortion lifts the degeneracy
of the $t_2$ manifold, yielding a doubly degenerate $e$ level and a
non-degenerate $a_1$ level. The $e$ states
accommodate three electrons, while the $a_1$ state remains unoccupied.

The calculated formation energy for the neutral vacancy
is 7.44~eV under C-rich conditions
and 7.99~eV under Si-rich conditions. The tetrahedral configuration remains
energetically competitive, with formation energies of 7.52~eV (C-rich) and
8.08~eV (Si-rich), in good agreement with the values reported in
Ref.~\cite{sc-sc21}. Previous {\em ab initio} studies have reported formation
energies ranging from 6.8 to 7.5~eV under C-rich conditions
\cite{sc-fa22,sc-zh24}, and from 7.3 to 8.3~eV under Si-rich conditions
\cite{sc-fa22,sc-zh24,sc-bo03,sc-sc21,sc-fa24}. We further note that the
formation energies of silicon vacancies in other technologically relevant
SiC polytypes, including 4H- and 6H-SiC, are comparable to those obtained
for the cubic phase \cite{sc-iw16,sc-ji22,sc-wa25}.

\begin{figure}
\vspace{-7mm}
\includegraphics[width=8cm]{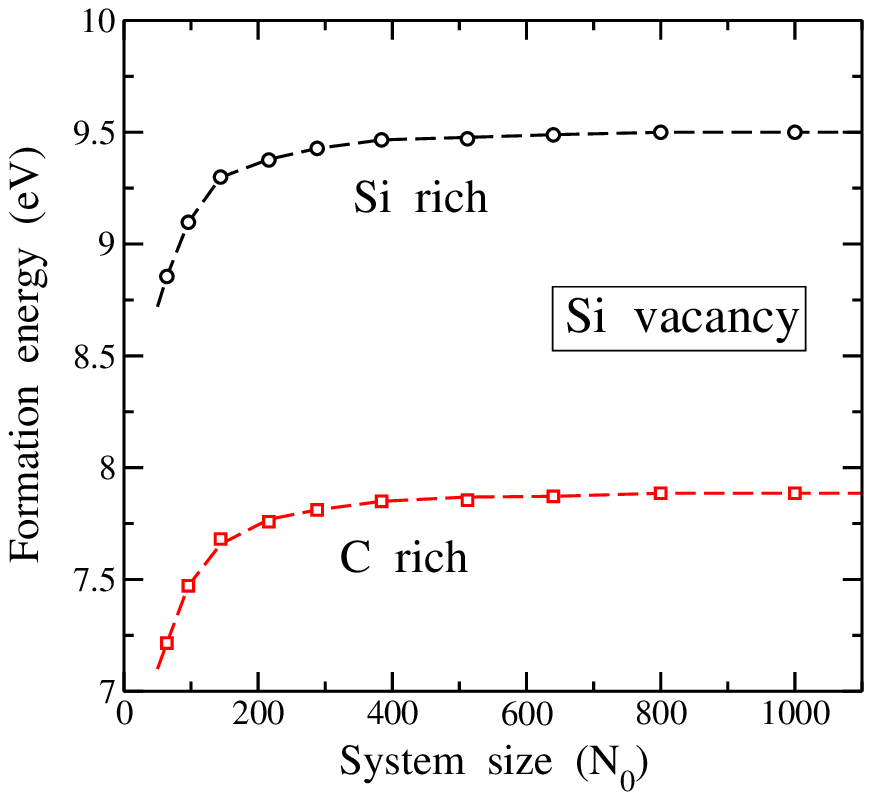}
\vspace{-5mm}
\caption{Formation energy of a silicon vacancy as a function of supercell
size for $3C$-SiC, calculated with the TB Hamiltonian used in this work.
Open circles and squares correspond to Si-rich and C-rich environments,
respectively.  Lines are guides to the eye.
}
\label{f1}
\end{figure}

The dependence of the calculated formation 
energy for the $V_{\rm Si}$ defect upon the supercell size has been 
analyzed earlier for various charge states \cite{sc-br11,sc-de18b,sc-sc21}.
Going to the results obtained with the TB method, we show in Fig.~1 
the formation energy $E_f$ of a neutral silicon vacancy in
$3C$-SiC as a function of the supercell size $N_0$.
The results are reported for Si-rich (circles) and C-rich (squares)
environments and were obtained from energy minimization calculations
(classical $T = 0$ limit) for each supercell size.
These data correspond to 
a vacancy concentration $x_v = 1 / N_0$, spanning the range from 
$10^{-3}$ to $1.6 \times 10^{-2}$.
To extrapolate the formation energy to the large-size limit
($N_0 \to \infty$), we fit the finite-size data shown in Fig.~1 
to the linear relation $E_f = E_f^{\infty} + c \, x_v$, where $c$
is a fitting parameter that accounts for residual size effects.
Restricting the fit to supercells with $N_0 > 140$, we obtain
$E_f^{\infty} = 7.93(2)$~eV and 9.54(2)~eV under C-rich and Si-rich
conditions, respectively. These formation energies, obtained within 
the TB model, are somewhat higher than the corresponding DFT values
reported above for the neutral vacancy under both chemical limits.

\begin{figure}
\vspace{-7mm}
\includegraphics[width=8cm]{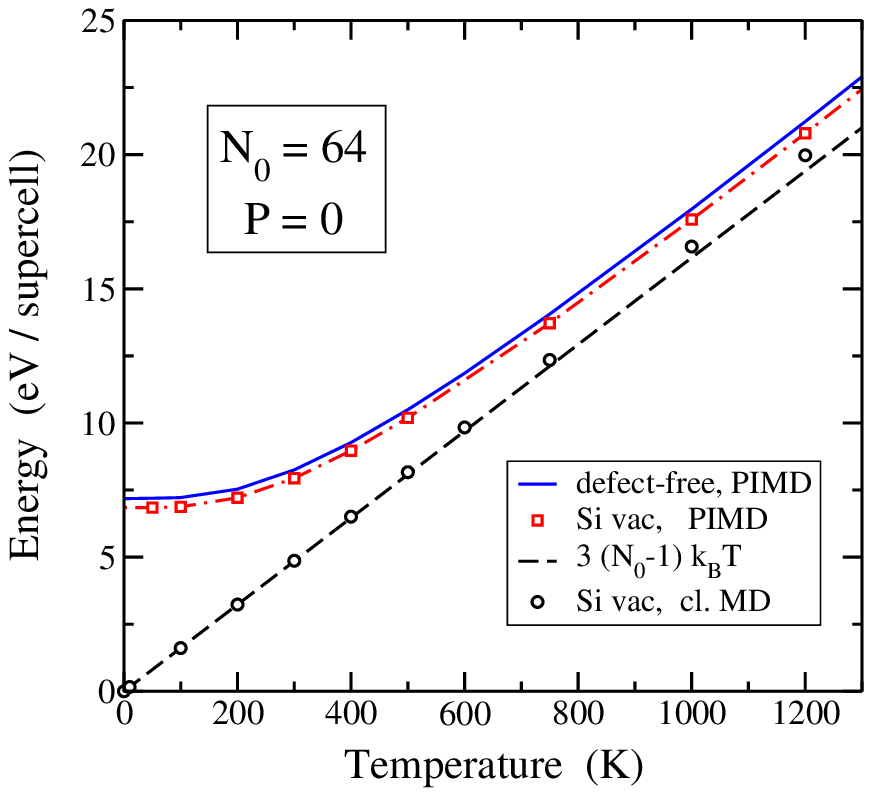}
\vspace{-5mm}
\caption{Energy $E - E_0$ vs temperature for a SiC supercell
($N_0 = 64$) containing a silicon vacancy.
Symbols display results of classical MD (open circles) and
PIMD simulations (open squares).
The label ``cl'' denotes ``classical''.
The dashed line represents the classical thermal energy:
$E_{\rm cl} = 3 (N_0 - 1) k_B T$, and the
dashed-dotted line through the PIMD results is a guide to
the eye. The solid curve represents the outcome of PIMD
simulations of defect-free $3C$-SiC with $N_0 = 64$
atoms, referred to its minimum-energy configuration
\cite{sc-he24}.
}
\label{f2}
\end{figure}

We now turn to our simulation results for SiC with silicon vacancies 
as a function of temperature $T$ and pressure $P$. 
The total energy is expressed as
$E = E_0 + E_{\rm pot} + E_{\rm kin}$, where $E_{\rm pot}$ and
$E_{\rm kin}$ denote the potential and kinetic energy, respectively, and
$E_0$ is the reference energy of the classical model at $T = 0$ and
$P = 0$, corresponding to the minimum-energy configuration.
In Fig.~2 we present the energy difference $E - E_0$ as a function of
temperature at vanishing pressure ($P = 0$).
Open circles and squares represent results from classical MD and PIMD
simulations, respectively, for a supercell with $N_0 = 64$.
The broken line shows the classical thermal energy,
$E_{\rm cl} = 3 (N_0 - 1) k_B T$.
At low temperatures, the classical MD data closely follow this linear
behavior, while at higher temperatures they progressively deviate from
the harmonic expectation, with clear departures becoming apparent in
Fig.~2 for $T > 700$~K.

The energies obtained from PIMD simulations are systematically higher than
their classical counterparts.
In the limit $T \to 0$, we obtain a zero-point energy $E_{ZP} = 6.84$~eV
(or $\overline{E}_{ZP} = 108$~meV/atom), and the two sets of data 
progressively converge as the temperature increases, reflecting 
the reduced importance of nuclear quantum
effects at higher $T$. This value of $\overline{E}_{ZP}$ is lower by 
4~meV/atom compared to that of perfect $3C$-SiC crystal \cite{sc-he24}.
Within the harmonic approximation, the zero-point energy of the 
defective supercell is given by
$E_{ZP} = 3 (N_0-1) \hbar \, \overline{\omega} / 2$, where 
$\overline{\omega}$ denotes the mean phonon frequency.
The reduction of $E_{ZP}$ in the presence of a vacancy therefore indicates
a softening of the atomic vibrational modes in the vicinity of the defect.
For a larger supercell with $N_0 = 216$, we obtain at low temperature
$\overline{E}_{ZP} = 111$~meV/atom, which is closer to the value reported 
for defect-free SiC (112~meV/atom \cite{sc-he24}).

\begin{figure}
\vspace{-7mm}
\includegraphics[width=8cm]{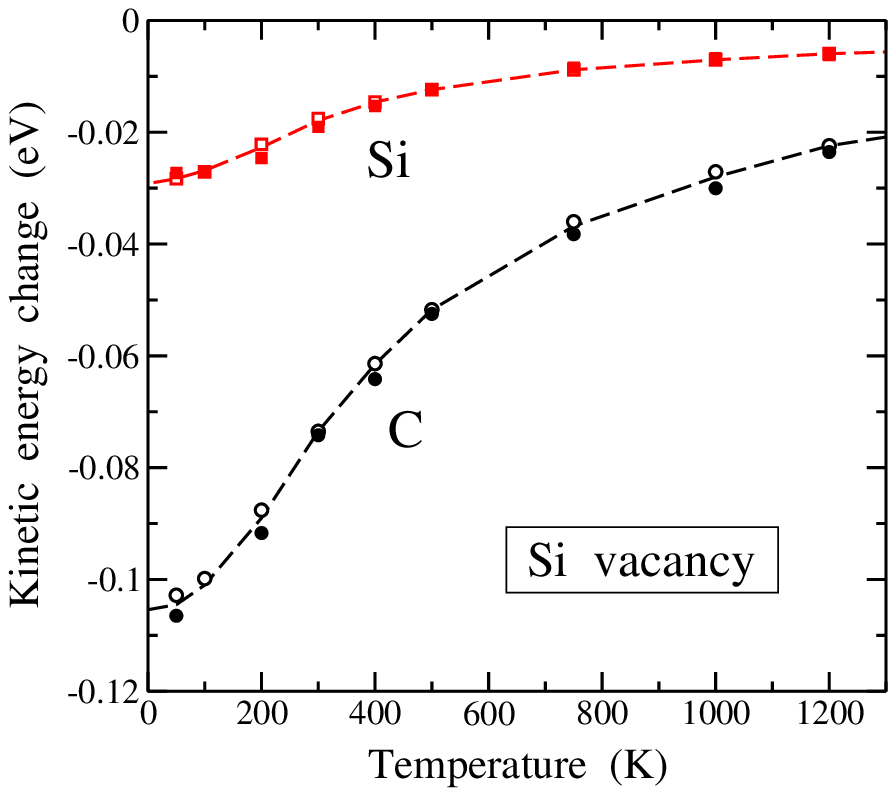}
\vspace{-5mm}
\caption{Temperature dependence of the change in overall kinetic
energy of C (circles) and Si atoms (squares), as obtained
from PIMD simulations for $3C$-SiC supercells with a single
Si vacancy. Open and solid symbols correspond to
$N_0 =$~64 and 216, respectively.
Error bars are on the order of the symbol size.
Dashed lines are guides to the eye.
}
\label{f3}
\end{figure}

The changes in vibrational frequencies induced by the vacancy are also 
reflected in the atomic kinetic energy at finite $T$. To quantify this 
effect, we calculate the difference between the total $E_{\rm kin}$ of 
C or Si atoms in a supercell containing a Si vacancy and that of the same 
number of atoms for the perfect SiC crystal.
In Fig.~3, we show the temperature dependence of the variation in the 
total kinetic energy of $N_0/2$ carbon atoms (circles) and $N_0/2 - 1$ 
silicon atoms (squares) in the supercell, resulting from the presence of 
the vacancy. Open and solid symbols correspond to supercells with 
$N_0 = 64$ and $216$, respectively.
In a classical model, this difference would vanish, since the kinetic 
energy per atom is given by 
$\overline{E}_{\rm kin}^{\rm cl} = 3 k_B T / 2$, 
independent of the local environment or interatomic interactions 
(equipartition theorem). At low temperatures, we find for $N_0 = 64$ 
reductions in 
$E_{\rm kin}$ of 105(1)~meV for C atoms and 29(1)~meV for Si atoms. 
The larger decrease for carbon arises from the stronger quantum 
delocalization of atoms near the vacancy.
The data in Fig.~3 also indicate a slight dependence on supercell size: 
the kinetic energy changes for carbon are marginally smaller for 
$N_0 = 216$ compared to $N_0 = 64$, likely due to a minor finite-size 
effect. The error bars are comparable to the symbol size.

Our simulations, both classical and PIMD, provide separate evaluations 
of the kinetic and potential energy of the solid \cite{ra11,he14,he22}. 
This distinction can be used to probe anharmonicities in the lattice 
vibrations by examining deviations between these energies, which would be 
identical in the purely harmonic limit. Anharmonic effects are most 
pronounced in the PIMD results, as they persist down to low temperatures 
due to zero-point motion.
For a Si vacancy, our simulations yield a ratio 
$E_{\rm kin} / E_{\rm pot} = 0.96$ at low $T$, decreasing to 0.92 at the 
highest temperatures considered. This corresponds to a difference between 
the kinetic and potential energies that grows from 4\% to 8\% with 
increasing temperature, highlighting the increasing importance 
of anharmonicity.

\begin{figure}
\vspace{-7mm}
\includegraphics[width=8cm]{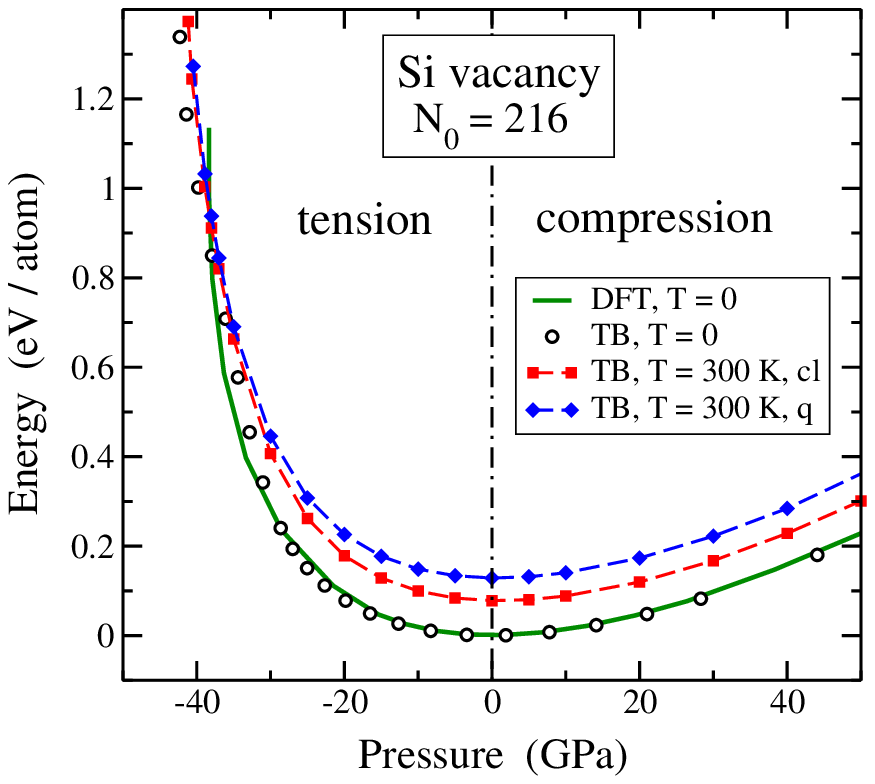}
\vspace{-5mm}
\caption{Energy per atom vs hydrostatic pressure $P$.
The solid line and open circles represent the results obtained
from DFT and TB calculations, respectively, at $T = 0$.
Solid symbols indicate data from classical MD (squares)
and PIMD simulations (diamonds) at $T = 300$~K.
The labels ``cl'' and ``q'' denote ``classical'' and ``quantum'',
respectively.
}
\label{f4}
\end{figure}

We now examine the evolution of the energy under hydrostatic pressure 
$P$. In Fig.~4, we present the energy difference 
$\overline{E} - \overline{E}_0$ as a function of $P$ for a supercell 
with $N_0 = 216$ and a Si vacancy ($x_v = 4.6 \times 10^{-3}$). 
The solid curve and open circles 
correspond to energies obtained at $T = 0$ from DFT and 
TB calculations, respectively. In both cases, the reference energy is 
taken for the unstressed solid ($P = 0$).
The two data sets are in close agreement over a wide pressure range, 
including the entire interval shown in Fig.~4. The largest deviations 
between both sets occur under tension ($P < -30$~GPa), 
where the TB energy is slightly higher than the DFT result. 
For pressures approaching $P \approx -36$~GPa, 
near the mechanical stability limit of the material (spinodal point, 
where the bulk modulus $B \to 0$; see below), this trend reverses, with 
the TB energy becoming lower than the DFT energy.

Solid symbols in Fig.~4 represent simulation results at $T = 300$~K. 
For the classical data, we observe an almost rigid upward 
shift of $3 k_B T$ per atom, relative to the tight-binding results at zero 
temperature. This indicates that anharmonic contributions to the classical 
system energy are minimal in the pressure range considered at $T = 300$~K.
At $P = 0$, the PIMD simulations show an energy increase 
$\delta \overline{E} = 51$~meV/atom relative to the classical result. 
This difference rises to 56~meV/atom under compression at 
$P = 60$~GPa and decreases to 
27~meV/atom under tension at $P = -35$~GPa. This behavior is primarily 
due to the increase of the mean phonon frequency $\overline{\omega}$ with 
hydrostatic pressure, which enhances the difference $\delta E$ between 
quantum and classical energies, particularly at relatively low temperatures.
At low $T$, the rate of change of $\delta E$ with pressure can be
expressed as:
\begin{equation}
 \frac {\partial (\delta \overline{E})} {\partial P} = 
  \frac {3 \hbar}{2} \, \frac {\partial \overline{\omega}} {\partial P} \; ,
\label{dde}
\end{equation}
an expression that is generally positive, as shown below.

The quantum zero-point expansion is governed by the anharmonicity
of the vibrational modes. Within the quasi-harmonic approximation,
each phonon mode contributes to the $T = 0$ crystal expansion through
the product of its zero-point energy and the corresponding
Gr\"uneisen parameter $\gamma_{\omega}$ \cite{mo05,de96,he20c}.
One may also introduce an overall, mode-independent Gr\"uneisen
parameter $\overline{\gamma}$, defined in terms of the mean frequency
$\overline{\omega}$ as \cite{sc-he24,as76}:
\begin{equation}
 \overline{\gamma} = - \frac {\partial ({\rm log} \, \overline{\omega})}
    {\partial ({\rm log} \, V) } = - \frac {V}{\overline{\omega}}
     \frac {\partial \overline{\omega}} {\partial V} \; .
\label{gamma}
\end{equation}
Using the definition of the isothermal bulk modulus,
$B = - V (\partial P / \partial V)_T$, Eq.~(\ref{dde}) can be 
rewritten as
\begin{equation}
\frac {\partial (\delta \overline{E})}{\partial P} = \frac {3 \hbar}{2} 
   \, \frac {\overline{\omega}}{B} \, \overline{\gamma} \; ,
\end{equation}
from which it follows that 
$\partial (\delta \overline{E}) / \partial P > 0$.
Indeed, $\overline{\omega}$ and $B$ are strictly positive, and
$\overline{\gamma}$ is generally positive as well; in particular,
for cubic SiC one finds $\overline{\gamma} \sim 1$
\cite{sc-va15,sc-zh13}.

We now compare the pressure dependence of the supercell energy in the
presence of a Si vacancy with that of the pristine crystal. In both cases,
the energies are referenced to their corresponding absolute minima,
i.e., $\Delta E = E - E_0$. At zero pressure and $T = 300$~K, the
classical value of $\Delta E$ for the perfect crystal is approximately
70~meV higher than for the defective supercell. This difference arises
mainly from the thermal contribution associated with the missing
Si atom in the defective system at 300~K.
In addition, a smaller residual contribution originates from the elastic
energy induced by the thermal lattice expansion at finite temperature.

Upon application of hydrostatic pressure, either compressive or tensile,
the energy of the defective crystal increases more rapidly than that of
the pristine material. Under compression, the difference between the
two systems grows steadily with increasing pressure, reaching
0.53~eV/supercell at $P = 50$~GPa. This behavior indicates that
the presence of the vacancy
enhances the energetic cost of compressive deformation, reflecting the
larger local lattice distortions induced around the defect site.
A similar trend is observed under tensile stress, although the effect
becomes substantially more pronounced. In the negative-pressure regime,
$\Delta E$ rises significantly faster for the defective supercell than
for the perfect crystal. For $P = -38$~GPa, which is close to the spinodal
instability discussed below, the energy difference reaches 8.6~eV/supercell.
The markedly stronger sensitivity of the vacancy-containing system to
tensile strain suggests that the defect weakens the mechanical stability
of the crystal under expansion, amplifying anharmonic lattice effects
and facilitating the onset of mechanical instability.
Note that this trend of the pressure-induced changes in $\Delta E$
corresponds to the classical limit. Considering both the defective
and perfect crystal in the quantum model, we find similar results,
since in both cases the main contribution to the energy change comes
from elastic deformations of the material, which are similar for
the quantum and classical cases.

\subsection{Atomic mean-square displacements}

We present results for the mean-square displacement (MSD),
$(\Delta {\bf r})^2 = \langle {\bf r}^2 \rangle -
\langle {\bf r} \rangle^2$, of C and Si atoms in 
vacancy-containing cubic silicon carbide, as obtained from 
our simulations.
PIMD provides a framework to calculate atomic MSDs as a function of 
temperature, allowing us to distinguish between a classical (thermal) 
contribution and an intrinsically quantum contribution.
Within the path-integral formalism, the classical part is associated 
with the motion of the center of gravity (centroid) of the quantum paths, 
whereas the quantum contribution is related to the mean spatial extent 
of the ring polymers that describe the quantum delocalization of 
the atomic nuclei.

\begin{figure}
\vspace{-7mm}
\includegraphics[width=8cm]{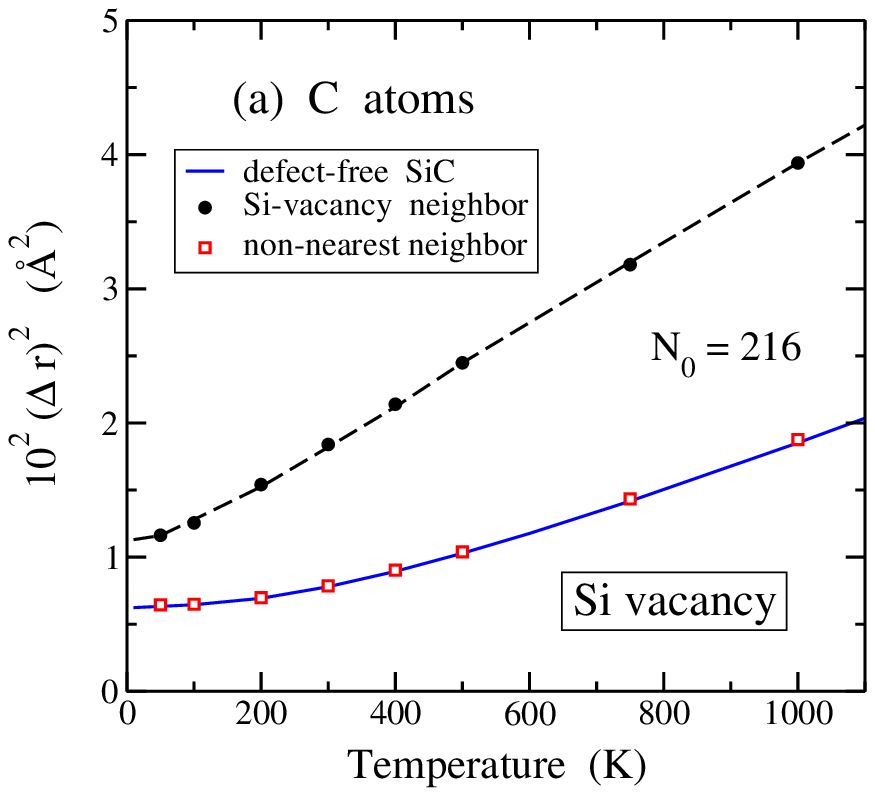}
\includegraphics[width=8cm]{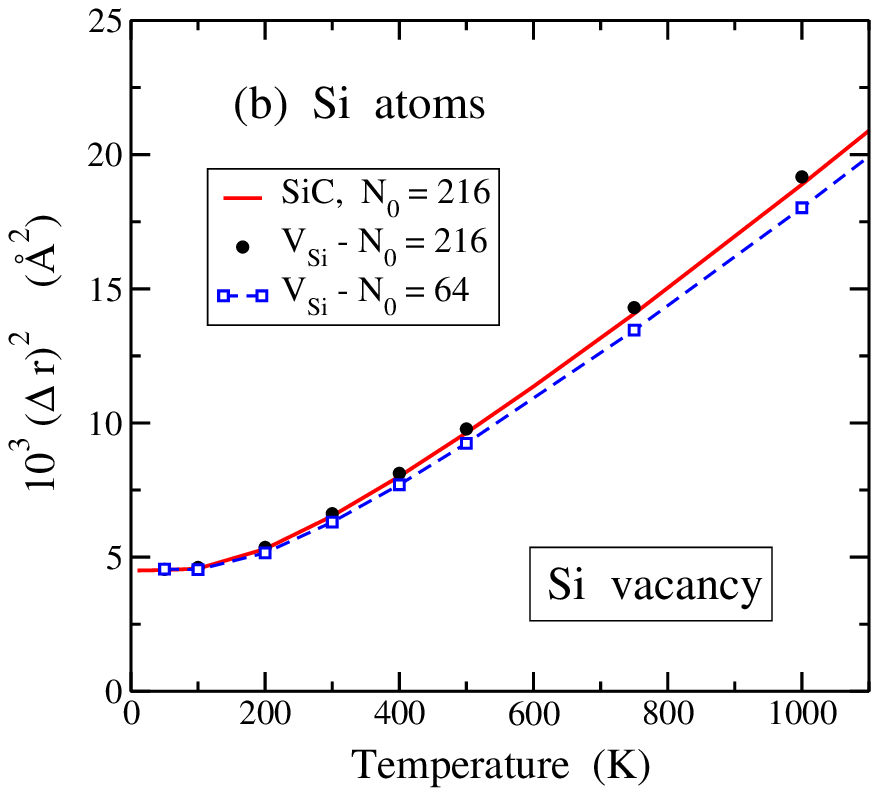}
\vspace{-5mm}
\caption{Atomic mean-square displacement, $(\Delta {\bf r})^2$,
in defective $3C$-SiC as a function of temperature for
(a) carbon and (b) silicon atoms.
Symbols represent results of PIMD simulations for a
vacancy-containing supercell with $N_0 = 216$.
Solid circles in (a) correspond to C atoms nearest-neighbor
of the vacancy, and open squares indicate the MSD of farther
atoms. Solid circles in (b) indicarte averages of MSDs for
all Si atoms in the defective supercell.
Open squares in (b) denote the MSD of silicon  atoms for
$N_0 = 64$ with an Si vacancy.
 Error bars are in the order of the symbol size.
Solid lines in (a) and (b) corresppond to MSDs in defect-free
cubic SiC.
Dashed lines through the simulation data are guides to the eye.
}
\label{f5}
\end{figure}

Fig.~5 shows the temperature dependence of the atomic MSDs derived from 
PIMD simulations.  Results are presented for (a) carbon and (b) silicon 
atoms in an SiC supercell with $N_0 = 216$.
For the C atoms, we find markedly different MSDs for nearest neighbors 
of the vacancy (solid circles) and for atoms farther away in the supercell 
(open squares), as expected from the additional free volume created 
by the missing Si atom at the vacancy.
In the low-$T$ limit, we obtain square displacements 
$(\Delta {\bf r})^2$ of $1.1 \times 10^{-2}$~\AA$^2$ and 
$6.4 \times 10^{-3}$~\AA$^2$, respectively, which correspond to zero-point 
motion.  The ratio between these two MSDs increases with temperature, 
from 1.7 at low $T$ to 2.0 at 1200~K.
For comparison, the solid line in Fig.~5(a) indicates the MSD of C atoms 
in the defect-free solid, which lies close to that of C atoms far from 
the vacancy in the defective supercell.

A similar trend is found for the MSD of Si atoms, shown in Fig.~5(b),
where atomic displacements in the defective crystal become slightly
larger than those in the perfect crystal with increasing temperature.
Open squares in Fig.~5(b) correspond to MSD values obtained for a
smaller supercell with $N_0 = 64$. At low temperature, these results
agree with those of the larger supercell within the error bars.
However, as the temperature increases, the MSD calculated for
$N_0 = 64$ becomes progressively smaller than that obtained for
$N_0 = 216$. This difference arises mainly from the contribution of
long-wavelength vibrational modes that are present in the larger
supercell but absent in the smaller one.
Indeed, for a cubic supercell of linear size $L$, the maximum
wavelength of the vibrational modes is effectively limited by
$\lambda_{\max} \approx L$, corresponding to a minimum wavenumber
$k_{\min} = 2 \pi / \lambda_{\max}$, which scales as
$k_{\min} \sim N_0^{-1/3}$. The influence of the supercell size is
most pronounced at high temperature, where quantum effects become
less important and atomic motion approaches the classical limit.
A detailed analysis of this finite-size effect for larger supercells
is beyond the scope of the present work, since a systematic convergence
study of finite-temperature MSDs remains computationally unaffordable
within our PIMD framework. In contrast, we have verified the
convergence of the lattice parameter $a$, as discussed below
in Sec.~III.C.

The atomic MSDs discussed here are closely related to the kinetic 
energy presented in Sec.~III.A.  In fact, a larger MSD corresponds 
to a lower kinetic energy $E_{\rm kin}$.
In the language of the ring-polymer representation used in PIMD, 
a larger radius of gyration is associated with a smaller $E_{\rm kin}$.
This relationship can be understood in terms of Heisenberg's uncertainty 
principle, as discussed below.

For a quantum particle of mass $M$, the root-mean-square deviations 
of a coordinate $x$ and its conjugate momentum $p_x$ satisfy the 
inequality $\Delta x \, \Delta p_x \geq \hbar / 2$ (see, {\em e.g.}, 
Ref.~\cite{co20a}).  Moreover, the kinetic energy,
$E_{\rm kin} = \langle {\bf p}^2 \rangle / 2M$, may be written as
\begin{equation}
     E_{\rm kin} = \frac{(\Delta {\bf p})^2}{2M} \; ,
\label{ekin}
\end{equation}
since in the absence of diffusion one has
$(\Delta {\bf p})^2 = \langle {\bf p}^2 \rangle -
\langle {\bf p} \rangle^2 = \langle {\bf p}^2 \rangle$,
because $\langle {\bf p} \rangle = 0$.
From the uncertainty principle, one obtains
\begin{equation}
    (\Delta p_x)^2 \geq \frac{\hbar^2}{4 (\Delta x)^2} \; .
\label{dpx2}
\end{equation}
Taking into account that, for a cubic crystal,
$(\Delta x)^2 = (\Delta y)^2 = (\Delta z)^2 = 
(\Delta {\bf r})^2 / 3$,
and combining Eqs.~(\ref{ekin}) and (\ref{dpx2}), we arrive at
\begin{equation}
   E_{\rm kin} \geq H \equiv 
	   \frac{9 \hbar^2}{8 M (\Delta {\bf r})^2 } \; ,
\label{ekinf}
\end{equation}
where $H$ is a function of the atomic MSD.

For a quantum particle in a given environment,
the function $H$ reaches its maximum value in the low-$T$ limit, 
where $(\Delta {\bf r})^2$ attains its minimum, corresponding to 
the zero-point MSD. Conversely, the kinetic energy $E_{\rm kin}$ 
approaches its minimum value as $T \to 0$.
Therefore, the zero-temperature limit provides a consistency 
check of our results through Eq.~(\ref{ekinf}).
Such consistency has previously been verified for perfect solids, 
in particular for silicon and carbon atoms in $3C$-SiC \cite{sc-he24}.
From the results shown in Figs.~3 and 5(a) for C atoms neighboring 
an Si vacancy, we obtain at low temperature
$E_{\rm kin} = 4.1 \times 10^{-2}$~eV
and $(\Delta {\bf r})^2 = 1.1 \times 10^{-2}$~\AA$^2$,
which yield a ratio $E_{\rm kin} / H = 1.15$.
The minimum possible value of this ratio, $E_{\rm kin} / H = 1$, is 
attained when $\Delta x \, \Delta p_x = \hbar / 2$ (and similarly for 
the other Cartesian coordinates), as realized for an isotropic 
three-dimensional harmonic oscillator in the limit $T \to 0$.
In solids, however, the presence of a dispersion of vibrational 
frequencies leads to values of $E_{\rm kin} / H$ that are 
systematically larger than unity.  For example, within a Debye 
model this ratio equals $9/8 = 1.125$ in the low-$T$ limit, 
independently of the Debye frequency $\omega_D$ \cite{he20b}.
This value is close to that obtained here for C atoms adjacent to 
a $V_{\rm Si}$ center, as well as for silicon and carbon atoms in 
defect-free $3C$-SiC \cite{sc-he24}.

The increased MSD of atoms neighboring the vacancy reflects a larger 
spatial extent of their quantum paths ({\em i.e.}, larger radius of 
gyration in the PIMD representation), which is a direct manifestation 
of zero-point motion in regions of reduced atomic coordination.
Thus, vacancies not only perturb the classical lattice but also amplify
the quantum fluctuations of nearby atoms. 
These enhanced quantum fluctuations near defects can have significant
consequences for material properties.
For example, the larger atomic delocalization may increase phonon scattering,
reducing thermal conductivity in defect-rich regions, and facilitate
defect-assisted diffusion by lowering the effective barriers for atomic 
migration.  Hence, understanding the MSD and quantum delocalization around 
vacancies is not only important for fundamental insight, but also for 
predicting the thermomechanical and transport behavior of silicon carbide 
and related materials.
Moreover, the quantum lattice fluctuations revealed by our simulations may 
have implications for defect-based spin qubits associated with the silicon 
vacancy in SiC. Since hyperfine couplings and spin-phonon interactions
depend sensitively on the local atomic configuration, nuclear quantum
delocalization could contribute to the renormalization of these interactions
and thereby affect spin-decoherence mechanisms.

\subsection{Volume}

For the minimum-energy configuration of the neutral silicon vacancy, 
the TB model employed in this work predicts a defect structure exhibiting 
a $C_{3v}$ distortion, in agreement with the DFT results (see Sec.~III.A). 
However, the $T = 0$ calculations indicate that the energy difference 
between the $C_{3v}$ ground state and the higher-symmetry $T_d$ 
configuration is very small, on the order of 10 meV. This near-degeneracy 
points to a shallow potential-energy landscape around the defect 
and suggests that a small thermal activation is required for the 
system to access symmetry-equivalent distorted configurations.
At finite temperatures, both classical MD and PIMD simulations yield 
a thermally averaged structure with apparent $T_d$ symmetry. Such behavior 
is expected when the symmetry-lowering distortion is associated with an 
energy scale comparable to thermal fluctuations. As the system evolves, 
it readily explores the equivalent $C_{3v}$ minima through thermally 
activated motion, and the resulting time-averaged atomic configuration 
therefore recovers the higher tetrahedral symmetry.
The effective $T_d$ symmetry observed in the finite-temperature simulations 
is maintained throughout the entire range of temperatures and hydrostatic 
pressures considered in this work.

Our DFT calculations at $T = 0$ indicate that the preferred
configuration of the $V_{\rm Si}$ center depends on the applied pressure. 
The silicon vacancy is known to be metastable with respect to the
$V_{\rm C}$-${\rm C_{Si}}$ complex, and at zero pressure both atom 
configurations are separated by an energy barrier of about 2~eV 
\cite{sc-ru04,sc-br11,sc-sc21}.
Under compression, the trigonal distortion associated to $V_{\rm Si}$
becomes more pronounced, as one of the carbon nearest neighbors to the 
vacancy progressively approaches the remaining three coplanar C atoms. 
For $P \gtrsim 55$~GPa ($a \approx 4.0$~\AA ), the $V_{\rm Si}$ center is 
no longer a minimum on the potential-energy surface and relaxes into 
the $V_{\rm C}$-${\rm C_{Si}}$ configuration.
Thus, our DFT results suggest that the energy barrier between both
configurations decreases with increasing hydrostatic pressure, 
although quantifying its pressure dependence 
requires further thorough analysis.
In any case, our DFT data indicate that the $V_{\rm Si}$ configuration 
becomes unstable at a pressure close to that where $3C$-SiC ceases to 
be the thermodynamically stable phase of the material ($P \sim 60$~GPa),
and therefore does not directly affect the present discussion, since
we focus on the effects of Si vacancies themselves on the
properties of the cubic phase.
Under tensile pressure, the tendency is opposite, as the
trigonal configuration progressively approaches a tetrahedral one.

We note that relatively small supercells may introduce interactions
between periodically repeated vacancies, leading to a finite dispersion
of defect-related states and influencing the local atomic relaxations.
In this context, employing different supercell sizes provides a direct
means of probing vacancy-concentration effects. In particular, the
$2 \times 2 \times 2$ ($N_0 = 64$) and $3 \times 3 \times 3$
($N_0 = 216$) supercells correspond to substantially different vacancy
concentrations, allowing us to identify systematic trends in the
calculated properties.
To assess the convergence of the equilibrium lattice parameter
predicted by the TB method at $T = 0$~K, we also considered systems
containing $N_0 =$~512 and 1000 sites, both for pristine and
vacancy-containing crystals. For these larger supercells, we obtain
the same value of $a_0$ for the perfect crystal within the precision
of our energy-minimization procedure. For supercells containing a
single vacancy, the calculated lattice parameters are
$a_0 =$ 4.3460~\AA\ and 4.3462~\AA\ for $N_0 =$ 512 and 1000,
respectively. These results, together with those obtained for smaller
supercells, indicate a linear dependence of $a_0$ on
the vacancy concentration $x_v$. A similar behavior was found in DFT
calculations of the lattice parameter at $T = 0$~K for $N_0 = 512$,
suggesting that the observed variations are primarily governed by the
vacancy concentration rather than by finite-size effects.
This issue is further discussed below in connection with the crystal
density at $T = 0$ and 300~K.

\begin{figure}
\vspace{-7mm}
\includegraphics[width=8cm]{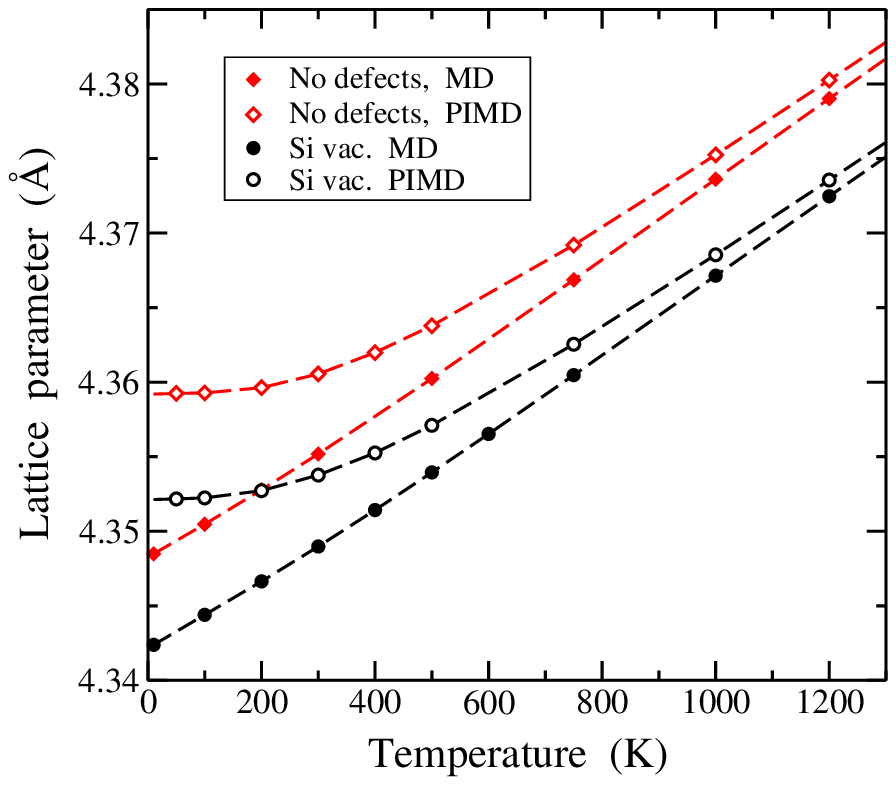}
\vspace{-5mm}
\caption{Temperature dependence of the lattice paramer $a$ of
$3C$-SiC with a single silicon vacancy ($x_v = 0.016$).
Solid and open circles represent results of classical
MD and PIMD, respectively.
For comparison, data for defect-free silicon carbide are shown as
solid diamonds (classical MD) and open diamonds (PIMD).
Dashed lines are guides to the eye.
}
\label{f6}
\end{figure}

We now analyze the effect of $V_{\rm Si}$ centers on the crystal 
volume of $3C$-SiC, as derived from TB data at finite temperatures.
In the following, we consider the vacancy 
concentration as $x_v = n_v / N_0$, where $n_v$ is the number
of Si vacancies in the supercell.
In Fig.~6 we show the temperature dependence of the lattice
parameter $a$ of unstressed $3C$-SiC, obtained for a simulation cell
with $N_0 = 64$, containing a single silicon vacancy ($x_v = 0.016$).
Symbols represent the results of our simulations:
solid circles correspond to classical MD, while open circles
denote PIMD calculations.
The lattice parameter $a$ obtained from classical simulations
exhibits an almost linear dependence on $T$, with a slope
$\partial a / \partial T$ that increases slightly with temperature.
At low temperatures, $a$ converges to a value $a_0 = 4.3421$~\AA,
with a slope $\partial a / \partial T = 2.2 \times 10^{-5}$~\AA/K.
Around $T = 1000$~K the slope increases to
$2.6 \times 10^{-5}$~\AA/K.
In PIMD simulations (open circles), we obtain a larger lattice
parameter, converging at low $T$ to $a = 4.352$~\AA.
This corresponds to an increase of 0.01~\AA\ due to zero-point
expansion with respect to the classical result.
Although this difference may appear small, it is much
larger than the precision limit of lattice parameters of
semiconductors that has been achieved for many years using
diffraction techniques \cite{ka98}.

For comparison with our results for defect-bearing SiC, Fig.~6 
also includes data obtained from both types of molecular dynamics
simulations for the ideal silicon carbide crystal with $N_0 = 64$
(diamonds).
In the presence of a Si vacancy, we observe a reduction of the
lattice parameter $a$ by $6 \times 10^{-3}$ and
$7 \times 10^{-3}$~\AA\ in the classical MD and PIMD results,
respectively. This means a defect-induced linear strain 
$\epsilon_L = \Delta a/ a = -1.4 \times 10^{-3}$ for 
the classical case and $-1.6 \times 10^{-3}$ for the quantum one.
This difference in lattice parameters decreases as the vacancy
concentration is reduced. For $N_0 = 216$
($x_v = 4.6 \times 10^{-3}$), it amounts to
$2 \times 10^{-3}$~\AA\ for both types of simulations, with
a defect-induced strain $\Delta a / a = -4.6 \times 10^{-4}$,
{\em i.e.}, about three times smaller than that for $N_0 = 64$.
We note that quantum and classical results for the lattice parameter
converge to each other as temperature is raised, for both defect-free
and vacancy-containing material, since in general 
quantum corrections become less relevant at high $T$.

\begin{figure}
\vspace{-7mm}
\includegraphics[width=8cm]{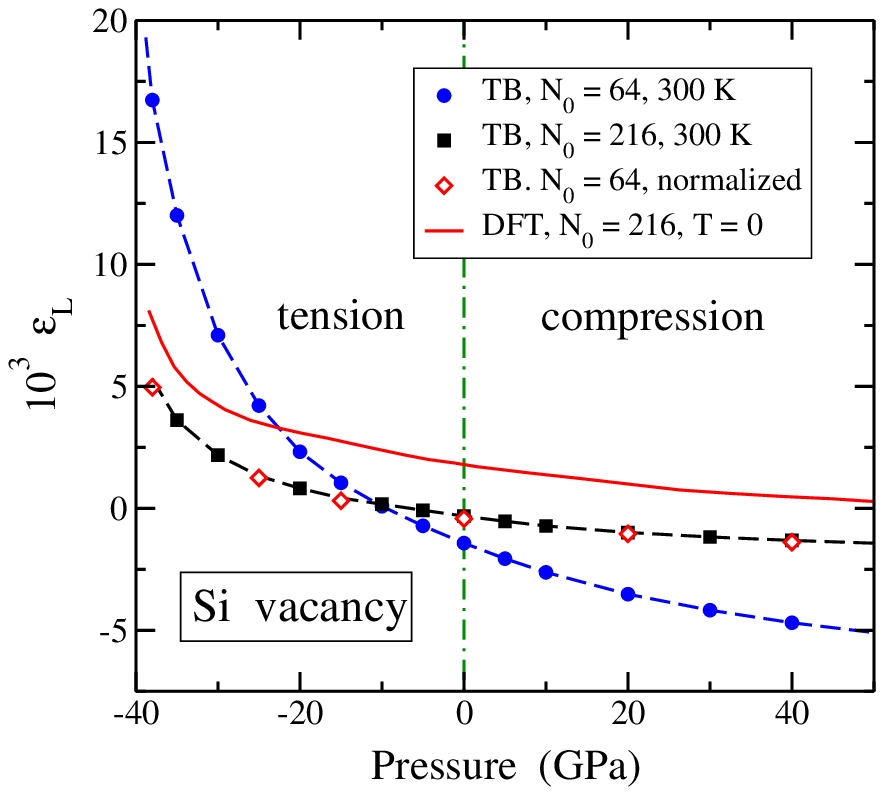}
\vspace{-5mm}
\caption{Pressure dependence of the strain
$\epsilon_L = \Delta a / a$ of silicon carbide with a Si
vacancy at 300~K. Solid circles and squares are data points
obtained from classical MD simulations for supercell size
$N_0$ = 64 and 216, respectively. Open diamonds represent the
strain for $N_0$ = 64 divided by the ratio 216/64 = 3.375.
Error bars are in the order of the symbol size.
The solid line corresponds to the DFT results for
$N_0 = 216$ at $T = 0$.
Dashed lines are guides to the eye.
}
\label{f7}
\end{figure}

The linear strain in vacancy-containing SiC exhibits a pronounced
dependence on applied pressure, with a particularly strong response
under tensile loading.
Fig.~7 shows the strain $\epsilon_L$ as a function of hydrostatic 
pressure obtained from classical MD simulations at $T = 300$~K:
solid circles for $N_0 = 64$ ($x_v = 0.016$) and solid squares for
$N_0 = 216$ ($x_v = 4.6 \times 10^{-3}$).
Under negative pressure (tension), the strain increases rapidly,
reaching values of 0.017 and $5 \times 10^{-3}$, respectively, 
at $P = -38$~GPa.
Upon further increase in tensile stress, the material becomes
mechanically unstable and approaches the spinodal point, at which
the pressure derivative of the volume (or, equivalently, of the
lattice parameter) becomes arbitrarily large, {\em i.e.},
$(\partial a / \partial P)_T \to -\infty$.
In contrast, under compressive pressure the strain decreases smoothly
with increasing $P$, attaining at $P = 50$~GPa values of
$-5.1 \times 10^{-3}$ and $-1.4 \times 10^{-3}$ for $N_0 = 64$
and 216, respectively.
A qualitatively similar pressure dependence of the strain is obtained
from PIMD simulations.  These results are systematically slightly 
larger than those obtained from classical MD, reflecting the influence 
of nuclear quantum effects.
However, the differences between the two datasets are small and
remain almost indistinguishable at the scale of Fig.~7.

One observes from our MD simulations that, for a given pressure $P$, 
the strain $\epsilon_L$ is proportional to the defect concentration $x_v$. 
In fact, the open diamonds in Fig.~7 correspond to the strain obtained 
for $N_0 = 64$ at several pressure values, divided by 3.375 (the ratio 
$216/64$). After this scaling, the data collapse onto the strain results 
for $N_0 = 216$, confirming the linear dependence on $x_v$.
An observable feature of the pressure dependence of the strain, shown 
in Fig.~7, is the sign change of $\epsilon_L$ as the applied pressure 
varies. Specifically, the strain is negative under compressive pressure 
and becomes positive under tensile pressure. For our TB results, this 
sign reversal is characterized by a relatively steep slope, 
$\partial \epsilon_L / \partial P = -1.3 \times 10^{-4}$~GPa$^{-1}$, for 
$N_0 = 64$. This behavior suggests that the pressure $P_c$, 
defined by $\epsilon_L = 0$, may depend sensitively on the particular 
computational model employed, as well as on the temperature $T$. For the 
representative case $T = 300$~K, we obtain $P_c = -9.9$~GPa, independently 
of the defect concentration $x_v$.

The solid line in Fig.~7 shows the strain $\epsilon_L$ obtained from 
our DFT calculations for $N_0 = 216$ at $T = 0$. Over the entire pressure 
range considered, the DFT values are approximately $2 \times 10^{-3}$ 
larger than the corresponding TB results for the same supercell size. 
A notable difference is that the DFT strain remains positive even at 
the highest compressive pressures investigated, whereas the TB values 
are negative for $P > 0$.
The most likely origin of this discrepancy lies in the limitations of 
the TB description under compression. Since the TB parametrization was 
optimized to reproduce structural and electronic properties near 
equilibrium, it may not fully capture the pressure dependence of the 
local bonding environment around the Si vacancy at positive pressures. 
While the TB model provides a reasonably accurate description of several 
pressure-dependent properties of the defective crystal, the quantity 
$\epsilon_L$ is determined by small differences in defect-induced 
lattice relaxations and is therefore particularly sensitive to 
inaccuracies in the underlying interatomic interactions.
We also note that zero-temperature TB calculations yield values of 
$\epsilon_L$ very similar to those obtained at $T = 300$~K and shown in 
Fig.~7. The discrepancy between the TB and DFT results is therefore not 
attributable to finite-temperature effects.

The pressure dependence of the strain, and in particular the fact
that $\partial \epsilon_L / \partial P < 0$, can be understood in
terms of the pressure-induced changes of the lattice parameter in
defect-free and vacancy-containing materials. For a cubic solid, one
generally has
\begin{equation}
   \frac{\partial a}{\partial P} = - \frac{a}{3 B} \; ,
\label{papb}
\end{equation}
where $B$ denotes the bulk modulus. The relative changes in the
lattice parameter $a$ caused by the presence of vacancies are smaller
than the corresponding changes in $B$ (see below), with the bulk
modulus being reduced in the defective material. Consequently, for a
given pressure $P$, the derivative in Eq.~(\ref{papb}) is more
negative for vacancy-containing SiC. This directly leads to a
negative pressure derivative of the strain.

\begin{figure}
\vspace{-7mm}
\includegraphics[width=8cm]{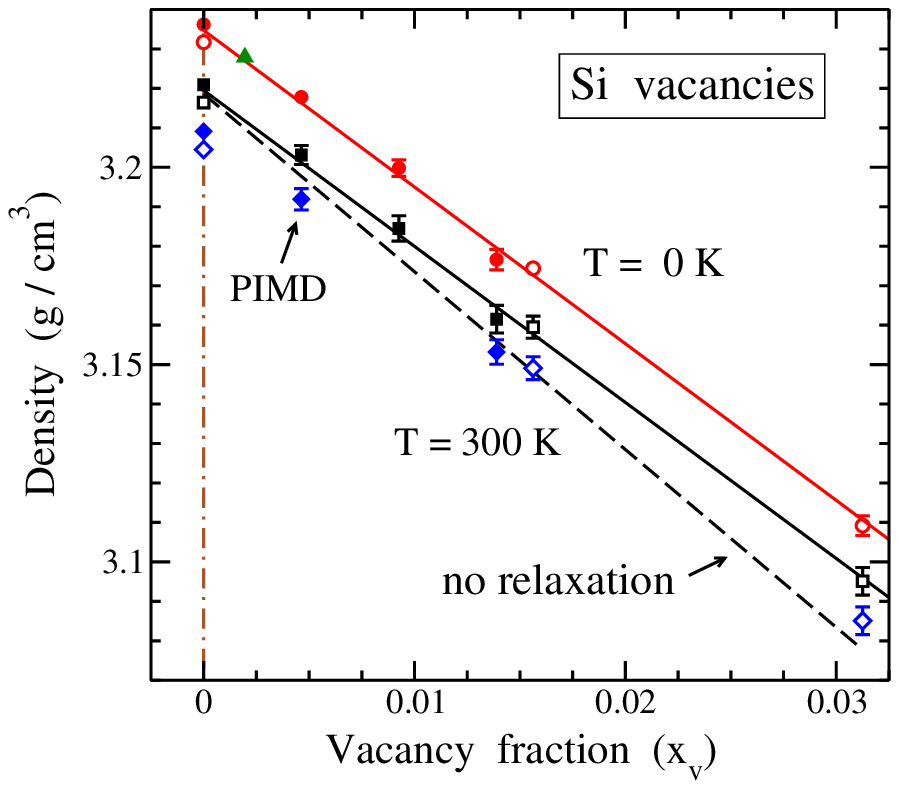}
\vspace{-5mm}
\caption{Density of $3C$-SiC vs fraction of silicon vacancies.
Data are given for $T = 0$ (energy minimization, circles) and 300~K
(classical MD simulations, squares; PIMD simulations, diamonds).
Open and solid symbols correspond to supercell size $N_0 = 64$
and 216, respectively. Solid lines are fits to the classical data.
A triangle indicates the density for $x_v = 1.9 \times 10^{-3}$
($N_0 = 512$) at $T = 0$~K.
Error bars, when not shown, are on the order or smaller than
the symbol size.
The broken line represents the expected density in the absence of
atomic relaxation around the vacancies at $T = 300$~K.
}
\label{f8}
\end{figure}

It is instructive to analyze the behavior of the crystal density 
$\rho$ in the presence of vacancies. 
On the one hand, the density is expected
to decrease due to the absence of Si atoms from their lattice sites;
on the other hand, this reduction may be partially compensated by
atomic relaxation in the vicinity of the vacancies.
In Fig.~8 we show the dependence of the material density on the
vacancy concentration $x_v$ for $P = 0$.
Circles represent the density at $T = 0$, as obtained from
energy minimization, while squares correspond to values derived
from classical MD simulations at $T = 300$~K.
In both cases, open and solid symbols denote supercells with
$N_0 = 64$ ($n_v = 1$ and 2) and $N_0 = 216$
($n_v = 1$, 2, and 3), respectively. The continuous lines are 
linear fits to the classical data at $T = 0$ and 300~K.
These lines are parallel over the range of vacancy concentrations
investigated here, with a slope
$\partial \rho / \partial x_v = -4.0$~g/cm$^3$ per vacancy/site.

It is worthwhile to comment on the origin of the error bars
associated with the data points in Fig.~8.
These uncertainties arise from two main sources.
First, for vacancy concentrations corresponding to more than one
defect per supercell ($n_v > 1$), the material density depends on
the relative spatial arrangement of the Si vacancies.
Second, there is the intrinsic statistical uncertainty associated
with volume fluctuations in the isothermal-isobaric simulations
performed at finite temperature.
To quantify the effect of vacancy arrangement, we considered five
different random configurations for each value of $x_v$.
The resulting dispersion in supercell volumes, together with the
statistical fluctuations, was used to estimate the error bars shown
in Fig.~8.  Overall, the results remain consistent with a linear 
dependence of the density on $x_v$, as discussed above.

In the absence of atomic relaxation in the vicinity of Si vacancies 
({\em i.e.}, assuming no change in volume), the density satisfies
\begin{equation}
   \frac{\rho(T)}{\rho_0(T)} = 1 - n_v \frac{M_{\rm Si}}{M_0} ; \,
\label{rhot}
\end{equation}
where $\rho_0(T)$ denotes the density of the ideal crystal at 
temperature $T$. The quantity
\begin{equation}
  M_0 = \frac{N_0}{2} (M_{\rm Si} + M_{\rm C})
\end{equation}
represents the total atomic mass of the defect-free supercell, 
with $M_{\rm Si}$ and $M_{\rm C}$ the atomic masses of Si and C, 
respectively.  Eq.~(\ref{rhot}) can be rewritten as
\begin{equation}
  \rho(T) = \rho_0(T) \left( 1 - x_v \frac{M_{\rm Si}}
	{\langle M \rangle} \right) \; ,
\label{rhot2}
\end{equation}
where $\langle M \rangle = M_0 / N_0$ is the average atomic mass.

The expected dependence of the density on $x_v$ at 300~K in the
absence of atomic relaxation is shown in Fig.~8 as a dashed line.
In this limit, one has
$\partial \rho / \partial x_v = - \rho_0 \, M_{\rm Si}/\langle M \rangle$,
which for $T = 300$~K yields a value of $-4.5$~g/cm$^3$ per 
vacancy/site.  This slope is somewhat more negative than that 
obtained from the simulations.
The discrepancy indicates that atomic relaxation in the vicinity of
the vacancies induces a contraction of the material ({\em i.e.}, an 
increase in density), which compensates approximately 11\% of the 
density reduction associated with the decreased mass of the supercell 
due to the presence of vacant sites.

Within the framework of density analysis, nuclear quantum motion,
or phonon quantization, is expected to reduce the density as a
consequence of the lattice expansion associated with such motion.
Open and solid diamonds in Fig.~8 denote the densities obtained from 
PIMD simulations for Si vacancies in supercells with $N_0 = 64$ and 
216, respectively, at 300~K. 
At this temperature, we observe a reduction in density of approximately 
0.01~g/cm$^{-3}$ relative to the corresponding classical results. 
This decrease is essentially independent of the vacancy concentration 
$x_v$ within the range considered.

\subsection{Elastic constants}

In this section, we analyze the effect of silicon vacancies on the
elastic constants of $3C$-SiC. The compliance elastic constants,
$S_{ij}$, are calculated at several temperatures by applying selected
components of the stress tensor $\sigma_{ij}$ in isothermal-isobaric
simulations. For example, when $\sigma_{xx} \neq 0$ and all other
stress components vanish ($\sigma_{ij} = 0$ for $ij \neq xx$), one
obtains $S_{11} = \epsilon_{xx} / \sigma_{xx}$ and
$S_{12} = \epsilon_{yy} / \sigma_{xx}$, where $\epsilon_{ij}$ are the
components of the strain tensor extracted from classical MD or PIMD
simulations \cite{as76,ki05,yu96}. To determine $S_{44}$, a shear
stress $\sigma_{xy}$ is applied, yielding
$S_{44} = \epsilon_{xy} / \sigma_{xy}$.
The stiffness constants $C_{11}$, $C_{12}$, and $C_{44}$ are then
obtained from the compliance constants using the standard relations
for cubic crystals \cite{as76,ki05}:
\begin{eqnarray}
   C_{11} & = & \frac {S_{11} + S_{12}}
        {(S_{11} - S_{12}) (S_{11} + 2 S_{12})}  \;  , \label{c11}   \\
  C_{12} & = & - \frac {S_{12}} {(S_{11} - S_{12}) (S_{11} + 2 S_{12})}
        \; ,  \label{c12}   \\
  C_{44} & = & \frac{1} {S_{44}}   \; .   \label{c44}
\end{eqnarray}

\begin{figure}
\vspace{-7mm}
\includegraphics[width=7.5cm]{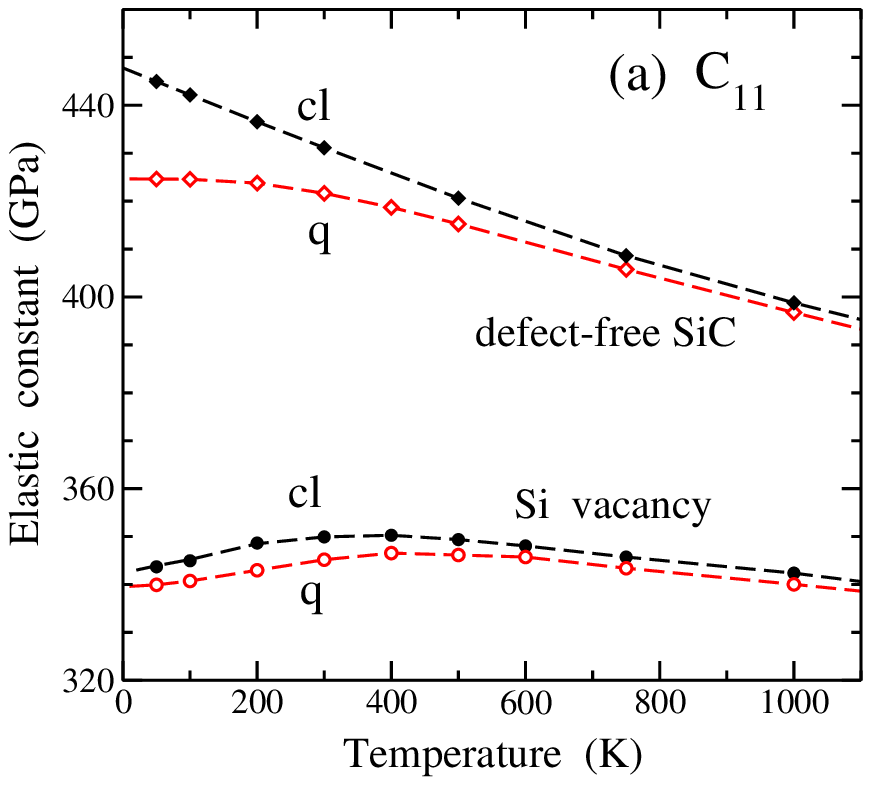}
\includegraphics[width=7.5cm]{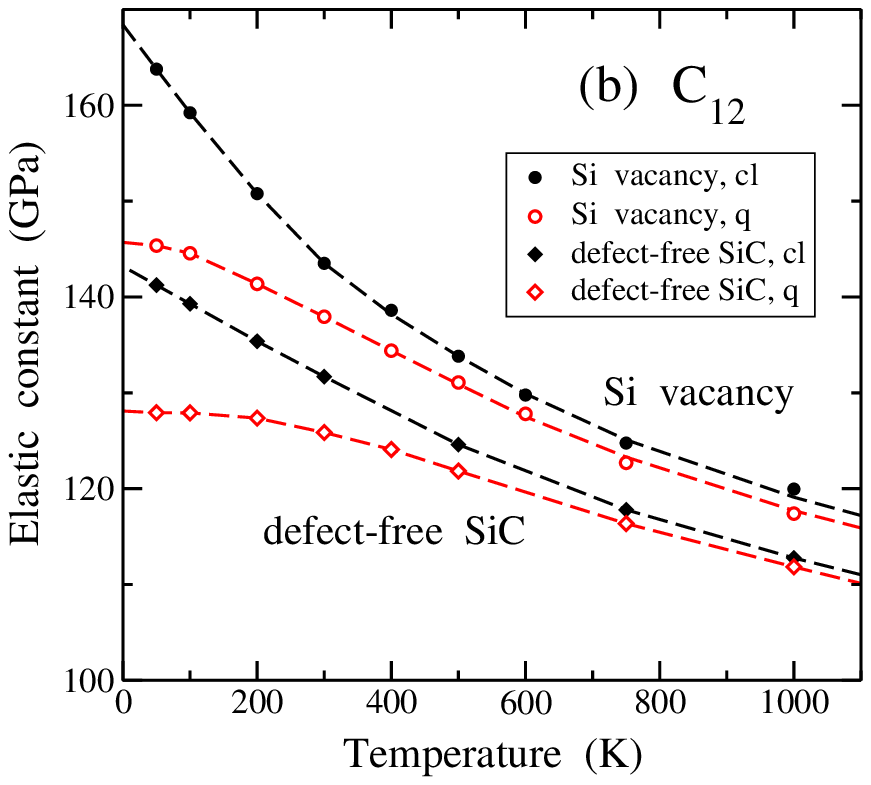}
\includegraphics[width=7.5cm]{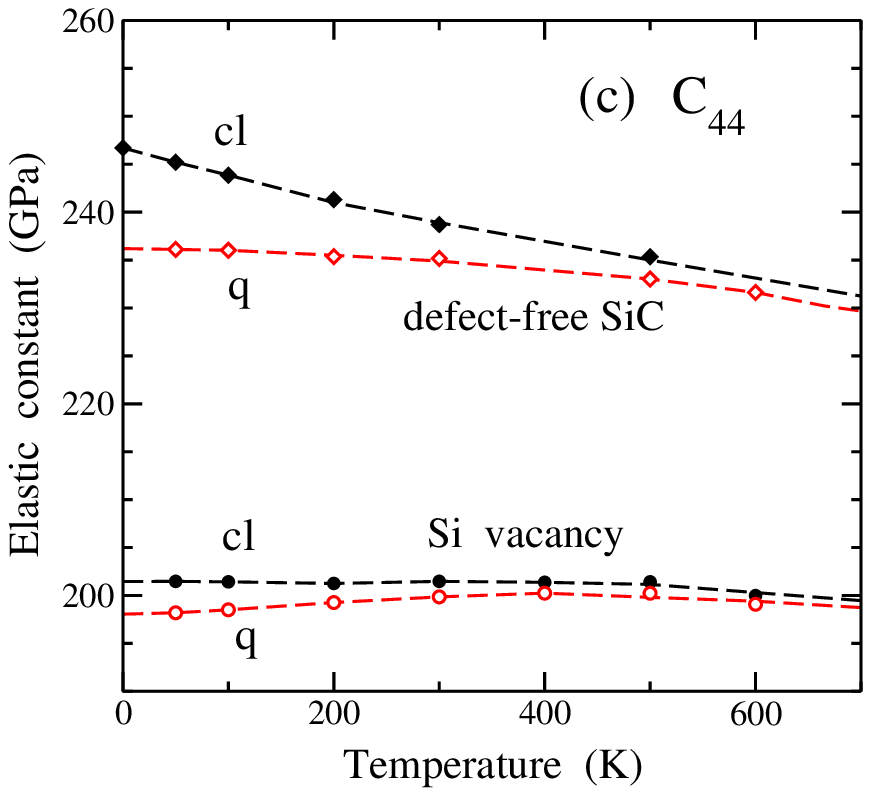}
\vspace{-5mm}
\caption{Temperature dependence of the stiffness elastic constants
of vacancy-containing $3C$-SiC:
(a) $C_{11}$, (b) $C_{12}$, (c) $C_{44}$, as derived from classical
MD (solid circles) and PIMD simulations (open circles).
For comparsison, solid and open diamonds represent outcomes of
MD and PIMD simulations for defect-free SiC, respectively.
The labels ``cl'' and ``q'' denote ``classical'' and ``quantum'',
respectively. Error bars are in the order of the symbol size.
Dashed lines are guides to the eye.
}
\label{f9}
\end{figure}

In Fig.~9, we show the temperature dependence of the stiffness
elastic constants of $3C$-SiC obtained from our simulations:
(a) $C_{11}$, (b) $C_{12}$, and (c) $C_{44}$.
Solid and open circles correspond to results for vacancy-containing
SiC ($x_v = 0.016$) derived from classical MD (labeled ``cl'') and 
PIMD simulations (labeled ``q''), respectively.
For comparison, the elastic constants of the perfect crystal are
also included in Fig.~9 and are represented by diamonds.
Considering first the results for the defect-free material, we find
that the classical elastic constants decrease with increasing
temperature, with slopes that become progressively less negative at
higher $T$. The quantum results for all three elastic constants are
systematically lower than their classical counterparts and approach
the zero-temperature limit with a vanishing slope
($\partial C_{ij} / \partial T \to 0$ as $T \to 0$).

Looking at the results for the vacancy-containing material 
(circles in Fig.~9), we observe that $C_{12}$ behaves similarly to 
the perfect crystal, following the trend seen in both classical and 
quantum data. In contrast, $C_{11}$ and $C_{44}$ exhibit an increase 
with temperature up to approximately 400~K, a behavior that is barely 
noticeable for the classical $C_{44}$ in Fig.~9(c). Overall, nuclear 
quantum effects reduce the stiffness constants of the defective solid, 
particularly at low temperatures, in a manner analogous to the 
ideal crystal.

Comparing the values of $C_{ij}$ for the defective and perfect solid, 
we find that the former exhibit an appreciable reduction in $C_{11}$ 
and $C_{44}$ due to the presence of Si vacancies, relative to the 
perfect crystal. Examining the low-temperature classical results for 
a defect concentration of $x_v = 0.016$, we observe decreases of 23\% 
and 18\% for $C_{11}$ and $C_{44}$, respectively. In contrast, $C_{12}$ 
increases in the presence of $V_{\rm Si}$ centers, rising by 17\% 
compared to the defect-free material in the classical low-$T$ limit.
A similar trend in the changes of the stiffness constants was reported 
by Fan {\em et al.} \cite{sc-fa22} based on DFT calculations at zero 
temperature for $x_v = 0.016$: they found decreases in $C_{11}$ and 
$C_{44}$ along with an increase in $C_{12}$ relative to the perfect 
crystal. The increase in $C_{12}$ in our results (17\%) is somewhat 
larger than that reported in those {\em ab initio} calculations (11\%).
The dependence of the stiffness constants on vacancy concentration in 
cubic SiC has also been investigated in Ref.~\cite{sc-ra24} using MD 
simulations. These authors did not observe a rise in $C_{12}$ 
for the vacancy-containing material, but rather a slight decrease.

It is instructive to analyze the relationship between the temperature 
dependence of the elastic constants and the atomic MSD discussed above. 
On the one hand, the classical results for the stiffness constants $C_{ij}$ 
display an approximately linear behavior at low $T$, associated with 
a linear increase in the MSD $(\Delta {\bf r})^2$ as temperature rises.
On the other hand, the low-$T$ elastic constants derived from PIMD 
simulations are reduced relative to the classical values due to nuclear 
quantum motion. In this case, the temperature derivative tends to zero 
in the zero-temperature limit, {\em i.e.}, 
$\partial C_{ij} / \partial T \to 0$ as $T \to 0$.
At higher temperatures, the classical and quantum results for $C_{ij}$ 
progressively converge, mirroring the analogous convergence observed in 
the MSD \cite{sc-he24}. A similar behavior is found for the bulk modulus 
in both classical and quantum simulations, as discussed below 
in Sec.~III.E.

In the context of elastic properties, the Poisson's ratio, $\nu$, 
characterizes the relationship between transverse and longitudinal 
strains under an applied stress. For cubic SiC, it is calculated as 
$\nu = -S_{12}/S_{11}$ \cite{sc-la91}. At $T \to 0$, our classical 
simulations of defective SiC yield $\nu = 0.33$, compared to 
$\nu = 0.30$ from low-temperature PIMD data. This indicates that 
zero-point quantum motion reduces the Poisson ratio by approximately 9\%.
As the temperature increases, $\nu$ decreases in both classical and 
quantum simulations. At $T = 1000$~K, the classical and quantum values 
converge to $\nu = 0.26$, with differences within error bars.

For comparison, we note that applying the same procedure to 
defect-free SiC at low temperature yields a Poisson's ratio of 
$\nu = 0.24$ \cite{sc-he24}, which is lower than the value obtained 
for the defective material. The increase in $\nu$ in the presence of 
$V_{\rm Si}$ defects is related to the decrease in $C_{11}$ and the 
increase in $C_{12}$ observed in Fig.~9 relative to the perfect solid. 
Indeed, using Eqs.~(\ref{c11}) and (\ref{c12}), one has 
$\nu = C_{12} / (C_{11} + C_{12})$.
A similar increase in $\nu$ for vacancy-containing SiC has been 
reported by Rabiee {\em et al.} \cite{sc-ra24} from MD simulations 
using the empirical Tersoff potential, as well as by 
Qin {\em et al.} \cite{sc-qi23} and Fan {\em et al.} \cite{sc-fa22} 
from DFT calculations.
The elastic constants and Poisson's ratio discussed here
correspond to a relatively high vacancy concentration ($x_v = 0.016$), 
which produces significant changes relative to the perfect solid. 
To first order, these corrections scale linearly with $x_v$, as 
observed for the material density in Fig.~8 and for the bulk modulus 
discussed below in Sec.~III.E.

\subsection{Bulk modulus}

In this section, we focus on the isothermal bulk modulus,
$B = - V (\partial P / \partial V)_T$, with particular emphasis on
its dependence on temperature, pressure, and vacancy concentration.
An alternative to evaluating $B$ through the volume derivative of
the pressure is provided by the fluctuation formula \cite{la80}:
\begin{equation}
   B = \frac{k_B T V}{(\Delta V)^2} \; ,
\label{b_fluc}
\end{equation}
where $V$ denotes the volume of the simulation cell and 
$(\Delta V)^2$ are the mean-square volume fluctuations. 
This expression is especially well suited for atomistic simulations 
performed in the isothermal-isobaric ensemble, as it avoids the 
need to compute numerical derivatives from data obtained at 
different volumes or pressures.

\begin{figure}
\vspace{-7mm}
\includegraphics[width=8cm]{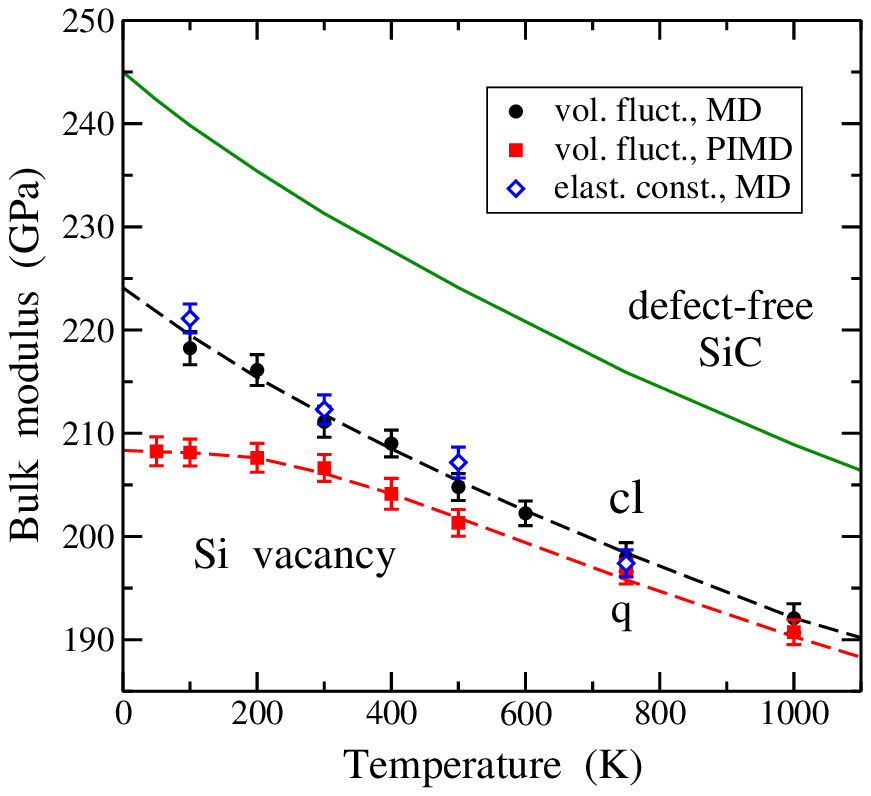}
\vspace{-5mm}
\caption{Temperature dependence of the bulk modulus of $3C$-SiC
with a single silicon vacancy ($x_v = 0.016$), obtained from the
volume fluctuations by using Eq.~(\ref{b_fluc}). Solid circles
and squares represent results drived from classical MD (denoted
as ``cl'') and PIMD simulations (denoted as ``q''), respectively.
Open diamonds indicate data for the classical bulk modulus calculated
from the elastic constants by means of Eq.(\ref{bulkm}).
The solid line represents the results for a perfect $3C$-SiC crystal
found from classical MD simulations \cite{sc-he24}.
Dashed lines are guides to the eye.
}
\label{f10}
\end{figure}

In Fig.~10 we show the temperature dependence of the bulk modulus $B$
for cubic SiC containing a silicon vacancy ($x_v = 0.016$), as
obtained from Eq.~(\ref{b_fluc}). Solid circles and squares denote
results from classical MD and PIMD simulations, respectively.
The classical data exhibit an almost linear decrease of $B$ at low
temperatures, characterized by a sizeable slope
$\partial B / \partial T$, which becomes progressively less negative
as the temperature increases. At $T = 1000$~K, $B$ is reduced by
approximately 15\% relative to its low-temperature value.
When nuclear quantum effects are included, $B$ is further reduced
with respect to the classical results, most noticeably at low
temperatures. In the limit $T \to 0$, this reduction amounts to
16(1)~GPa, corresponding to a 7\% decrease of the bulk modulus due to
atomic zero-point motion. As the temperature rises, the two data
sets converge, and their difference falls below 1\% at $T = 1000$~K.

The solid curve in Fig.~10 represents the temperature dependence of
the bulk modulus of defect-free $3C$-SiC, obtained from classical MD
simulations using the same TB approach. By comparing this curve with
the classical results for the vacancy-containing system, we observe
a systematic reduction of $B$ over the entire temperature range due
to the presence of a silicon vacancy. This decrease amounts to
20(1)~GPa at low temperatures and remains sizable, 16(1)~GPa, at
$T = 1000$~K.

To elucidate the low-temperature behavior of $B$, as obtained from
both classical MD and PIMD simulations using Eq.~(\ref{b_fluc}), 
we recall that the mean-square volume fluctuations scale as
$(\Delta V)^2 \sim T$. That is, at low temperatures they increase
linearly with $T$ in both classical and quantum treatments
\cite{sc-he24}. Consequently, the temperature dependence of $B(T)$ is
largely governed by the behavior of the volume $V(T)$. In the
classical limit, $V$ increases linearly with $T$, whereas in the
quantum case one finds $\partial V / \partial T \to 0$ for
$T \to 0$, as required by the third law of thermodynamics (vanishing
thermal expansion). According to Eq.~(\ref{b_fluc}), this implies, to
first order, a linear dependence of $B$ on $T$ in classical
simulations, while $\partial B / \partial T \to 0$ for the PIMD
results, in agreement with the third law \cite{ca85,be00b}. 
We note that this thermodynamic requirement is not
fulfilled by the classical data, for which $\partial B / \partial T <
0$ persists down to the lowest temperatures.
Finally, inspection of Eq.~(\ref{b_fluc}) together with the bulk
modulus obtained from our simulations shows that the reduction of
$B$ due to nuclear quantum effects at low temperatures is 
associated with an increase in the volume fluctuations
$(\Delta V)^2$ relative to the classical case. This increase is
proportionally larger than the corresponding increase in the
average volume $V$.

The isothermal bulk modulus can also be obtained from the elastic
constants through an expression valid for cubic crystals
\cite{as76,ki05,sc-ja14}:
\begin{equation}
    B = \frac{C_{11} + 2 \, C_{12}}{3} \; .
\label{bulkm}
\end{equation}
Open diamonds in Fig.~10 denote classical values of $B$ calculated
using Eq.~(\ref{bulkm}). These results are in close agreement with
those derived from the fluctuation formula in Eq.~(\ref{b_fluc}), 
providing a consistency check for our calculations.

\begin{figure}
\vspace{-7mm}
\includegraphics[width=8cm]{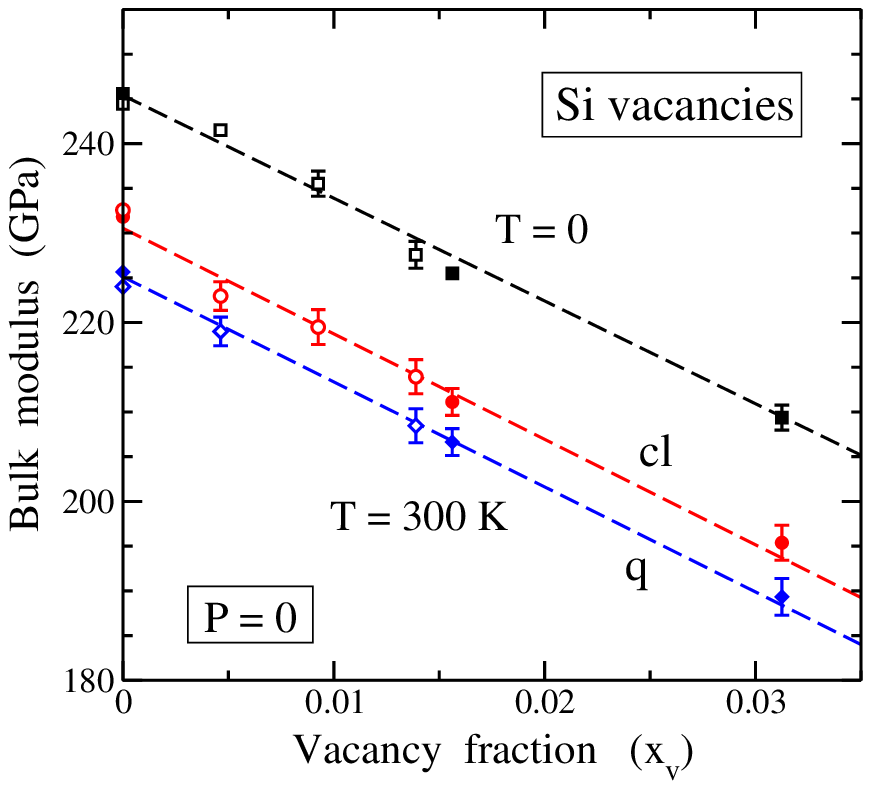}
\vspace{-5mm}
\caption{Bulk modulus of $3C$-SiC vs fraction of silicon vacancies.
Data are given for $T = 0$ (energy minimization) and at $T =$~300~K
for classical MD (circles, denoted as ``cl'') and PIMD simulations
(diamonds, denoted as ``q'').
Solid and open symbols are data points for supercell size
$N_0 = 64$ and 216, respecively.
Error bars, when not shown, are on the order or smaller than
the symbol size.
Dashed lines are fits to the data points.
}
\label{f11}
\end{figure}

We now turn to the dependence of $B$ on the vacancy concentration
$x_v$ in silicon carbide. Fig.~11 displays the bulk modulus as a
function of $x_v$ for $T = 0$ (minimum-energy configurations,
squares) and for $T = 300$~K (classical MD simulations, circles).
Results are shown for supercells with $N_0 =$~64 (solid symbols) 
and 216 (open symbols). Dashed lines represent linear fits to the 
data of the form $B = B(0) + b \, x_v$, where $B(0)$ is the
bulk modulus of the perfect crystal. The fitted parameter $b$ is
1155(50) and 1190(50)~GPa per vacancy/site for $T = 0$ and 300~K,
respectively. Within the uncertainty of both the data points and
the fitted parameters, the two $B(x_v)$ lines are nearly parallel.
As a consequence, for a vacancy concentration $x_v = 0.01$ we find a
reduction of the bulk modulus of approximately 12~GPa, corresponding
to about 5\% of $B(0)$.  
A further reduction in $B$ is found from PIMD simulations of defective 
silicon carbide at $T = 300$~K, as shown in Fig.~11 for various vacancy 
concentrations $x_v$ (diamonds). The bulk modulus decreases monotonically 
with increasing $x_v$, following the same qualitative trend observed in 
the classical simulations. Quantum nuclear fluctuations additionally soften 
the material, yielding lower values of $B$ throughout the studied 
concentration range. The dependence on vacancy concentration remains 
approximately linear and is characterized by a slope parameter 
$b = 1140(60)$~GPa per vacancy/site, similar to the values reported above.

For comparison, a linear fit to the DFT results for $B(x_v)$ yields
a slope parameter of $b = 890$~GPa per vacancy/site, somewhat smaller 
than the corresponding TB value. The significance of this difference is, 
however, limited by the small number of vacancy concentrations that 
can be explored with DFT calculations. As a consequence, it is difficult 
to assign a reliable uncertainty to the DFT-derived value of $b$ and 
to assess quantitatively the discrepancy between both approaches.
Nevertheless, the lower DFT value can be plausibly attributed to its 
more accurate treatment of the local atomic relaxations around vacancies. 
These relaxations tend to redistribute and partially delocalize the strain 
field induced by the defects, thereby reducing their impact on the 
macroscopic elastic response. In contrast, the more localized bonding 
description inherent to the TB model may enhance the apparent stiffness 
reduction associated with each vacancy, leading to a larger value of $b$.

As discussed previously for the crystal density (Sec.~III.C and Fig.~8),
the bulk modulus $B$ at a given defect fraction $x_v$ may also depend on
the specific spatial arrangement of the Si vacancies in configurations
with $n_v > 1$. Although our results do not reveal a strong sensitivity
to the defect distribution, a measurable dependence on the vacancy
arrangement is nevertheless observed. This effect contributes
additional variability beyond the statistical fluctuations associated
with the isothermal-isobaric ensemble employed in the simulations,
thereby increasing the resulting error bars.

At this point, we note that the Si vacancy concentrations realized in
our supercell calculations are necessarily higher than those typically
encountered in experimental samples.
This is a general limitation of finite-temperature defect simulations
with TB-derived interactions, where computational cost restricts
the accessible supercell sizes.
As a consequence, residual interactions between periodically repeated
vacancies may increase the magnitude of the calculated defect-induced
modifications, particularly for elastic properties. Nevertheless,
the structural perturbation associated with an isolated Si vacancy remains
relatively localized, suggesting that the trends obtained here
are representative of the dilute-defect regime. In the low-concentration
limit, the variation of elastic constants is expected to scale approximately
linearly with vacancy concentration. The present results may therefore be
regarded as finite-concentration reference values that can be extrapolated
toward experimentally relevant concentrations. This interpretation is
further supported by the nearly linear behavior obtained in our calculations
where the relative changes in the elastic response remain proportional 
to $x_v$ within the statistical uncertainty of the simulations. 
Such behavior indicates that,
in this regime, collective vacancy-vacancy effects are still limited and
the calculated elastic softening is primarily governed by the local
perturbation introduced by individual defects.

The linear trend obtained above for $B$ as a function of
$x_v$ can be used for an estimation of vacancy-induced changes
in the bulk modulus under experimentally accessible conditions, where
the defect concentration is expected to remain in the dilute regime.
Thus, our result $\partial B / \partial x_v \approx 1200$~GPa 
per vacancy/site defines a quantitative reference for such estimations. 
This suggests that Si vacancies may contribute 
to the elastic response of SiC in irradiated or nonstoichiometric samples,
particularly in situations where defect accumulation occurs during growth,
implantation, or prolonged operation under extreme conditions.
This discussion concerning the extrapolation of bulk-modulus variations
toward the dilute-defect limit can also be extended to other physical
properties of the crystal, such as the density $\rho$ shown in Fig.~8. 
In this case, the calculated dependence on vacancy concentration is 
likewise approximately linear within the explored range (see Sec.~III.C). 

\begin{figure}
\vspace{-7mm}
\includegraphics[width=8cm]{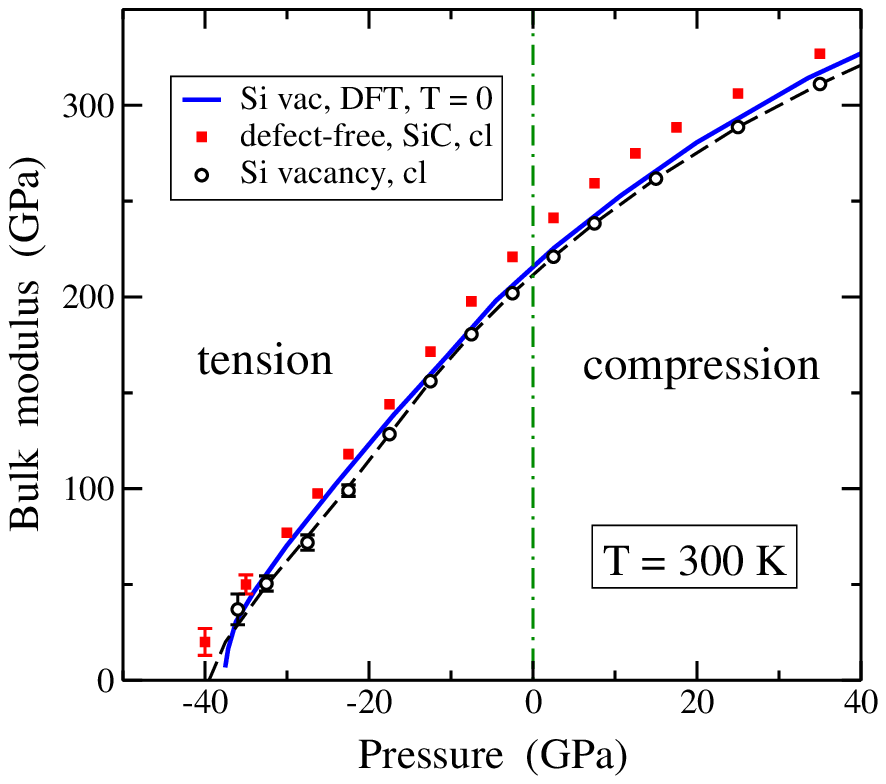}
\vspace{-5mm}
\caption{Bulk modulus of defective $3C$-SiC ($x_v = 0.016$), as
derived from classical MD at $T$ = 300~K for several hydrostatic
pressures (open circles).
The dashed line through the data points is a guide to the eye.
Solid squares represent the bulk moduluus of perfect $3C$-SiC,
obtained from classical MD simulations \cite{sc-he24}.
Results of DFT calculations at $T = 0$ are presented as
a continuous curve.
}
\label{f12}
\end{figure}

We now examine the dependence of the bulk modulus of vacancy-containing
$3C$-SiC on the hydrostatic pressure $P$. The results are shown in
Fig.~12, where both tensile ($P < 0$) and compressive ($P > 0$) regimes
are considered. Open circles correspond to classical MD simulations
performed at $T = 300$~K for a vacancy concentration $x_v = 0.016$.
Results from PIMD simulations at this temperature are almost
indistinguishable from the classical ones at the scale of the figure
and are therefore not shown.
The TB results at $T = 0$ are slightly larger than the classical values
at $T = 300$~K and are omitted for clarity. For comparison, we include 
the pressure dependence of the bulk modulus for the ideal crystal,
obtained from classical MD simulations (solid squares) 
at 300~K \cite{sc-he24}.
As expected, these values are systematically higher than those for
vacancy-containing SiC over the entire pressure range considered.
The solid line in Fig.~12 represents the evolution of $B$ with pressure,
as derived from our $T = 0$ DFT calculations, and lies close to the TB
results for the defective material at 300~K.

As in the defect-free crystal, vacancy-containing SiC exhibits a rapid 
decrease of $B$ with increasing tensile pressure, 
eventually vanishing at pressures close to $P \approx -40$~GPa. 
This behavior signals the onset of mechanical instability. 
Specifically, for $x_v = 0.016$ at 300~K, we obtain a spinodal pressure 
of $P_s = -39(1)$~GPa, compared to $P_s = -43$~GPa for the perfect 
crystal ($x_v = 0$) \cite{sc-he23}. Thus, the presence of vacancies 
reduces the stability region by approximately 4~GPa at this defect 
concentration. In the vicinity of the spinodal pressure, the bulk 
modulus follows the expected scaling law $B \sim (P - P_s)^{1/2}$, 
with a diverging derivative $\partial B / \partial P$ at $P_s$, 
in agreement with thermodynamic predictions \cite{sc-he23}.
In Sec.~III.A we assumed a spinodal pressure of $P_s = -39$~GPa 
for a vacancy concentration of $x_v = 0.016$ at 300~K, consistent 
with the divergence observed in the slope of the energy-pressure 
curve. This value of $P_s$ agrees with the extrapolation of the bulk 
modulus $B$ to zero under tensile conditions at this temperature.

In the present context, the term ``spinodal'' does not refer to phase 
separation in the classical thermodynamic sense (as in a binary 
mixture driven by compositional fluctuations). Rather, it denotes 
the mechanical stability limit associated with the loss of convexity 
of the free energy with respect to volume or strain, {\em i.e.}, 
the point at which $\partial^2 F / \partial V^2 = 0$
($F$ being the Helmholtz free energy \cite{ca85}). 
At this point, the bulk modulus $B$ vanishes, marking the emergence 
of mechanical instability. Consequently, 
at the spinodal pressure the solid can no longer sustain increasing 
tensile stress and becomes unstable with respect to strain fluctuations. 
In our atomistic simulations, the vicinity of the spinodal pressure 
$P_s$ is characterized by pronounced volume fluctuations, which may 
lead to crystal breakdown and, ultimately, to fragmentation.

This behavior is related to the calculations of the {\em tensile strength} of
defective $3C$-SiC reported by Li and Xiao \cite{sc-li19}, who performed MD
simulations of large supercells subjected to uniaxial stress. The tensile
strength measures the maximum stress (or strain) that a material can sustain
before mechanical failure. For the defect-free material, these authors found
a tensile strength of about 90~GPa, close to the stability limit under 
uniaxial pressure obtained in our previous study \cite{sc-he25}.
For silicon carbide containing $V_{\rm Si}$ defects, they observed a
significant decrease in tensile strength with increasing vacancy
concentration $x_v$. For $x_v = 0.016$, we estimate from their results a
reduction of the tensile strength of approximately 10~GPa. Dividing this
value by a factor of 3 to obtain the corresponding hydrostatic pressure
component associated to the uniaxial stress allows a direct comparison 
with the shift of 4~GPa found in our calculations of $P_s$, showing 
reasonable agreement between both results.

\section{Summary}

This paper presents a theoretical investigation of the structural and
elastic properties of $3C$-SiC containing silicon vacancies, with 
particular emphasis on the role of nuclear quantum effects and the 
mechanical stability of the material under tensile stress. Silicon carbide 
is widely used in electronic and structural applications, and Si vacancies 
have recently gained attention due to their relevance for quantum technologies. 
Understanding how these defects influence the mechanical response of SiC 
is therefore of both fundamental and technological importance.

The study is based on classical and path-integral molecular dynamics 
simulations employing a validated tight-binding Hamiltonian. This combined 
approach allows for a consistent comparison between classical behavior and 
quantum nuclear motion arising from zero-point fluctuations. Structural and 
elastic properties are analyzed over a wide range of temperatures and 
hydrostatic pressures, including both compressive and tensile regimes.

The presence of Si vacancies is found to significantly modify the elastic 
response of $3C$-SiC. In particular, the elastic constants $C_{11}$
and $C_{44}$, as well as the bulk modulus, are reduced compared to those 
of the defect-free crystal, and an increase is found for $C_{12}$. 
These changes reflect the local lattice distortions 
induced by vacancies and the resulting weakening of interatomic bonding. 
Nuclear quantum motion further causes reductions in all cases, especially 
at low temperatures, where zero-point motion leads to additional 
lattice softening.

A central result of this work concerns the mechanical stability of 
vacancy-containing $3C$-SiC under tensile pressure. The bulk modulus decreases 
rapidly with increasing tension and vanishes at a spinodal pressure that 
defines the limit of mechanical stability: $P_s = -39$~GPa for $x_v = 0.016$
at 300~K. The presence of Si vacancies 
shifts this spinodal pressure toward less negative values, reducing the 
stability region of the material. For the vacancy concentrations considered, 
the maximum sustainable tensile pressure is lowered by several gigapascals 
compared to the ideal crystal. Near the stability limit, large volume 
fluctuations are observed, indicating the onset of mechanical instability 
and eventual structural breakdown.

In summary, we have shown that silicon vacancies induce appreciable
modifications of the elastic constants and bulk modulus of cubic $3C$-SiC,
and that nuclear quantum motion further softens these properties, especially
at low temperatures. Vacancies also reduce the maximum sustainable tensile
pressure, altering the mechanical stability domain of the crystal.
These findings provide a quantitative framework for understanding the
interplay between point defects and lattice mechanics in SiC.
Moreover, the quantum-induced lattice fluctuations identified here may
have implications for defect-based spin qubits: since hyperfine couplings
and spin-phonon interactions are sensitive to the local atomic environment,
nuclear quantum delocalization could influence spin relaxation and
decoherence mechanisms in silicon-vacancy centers.

\begin{acknowledgments}
R. Ram\'irez is gratefully acknowledged for insightful discussions on 
the physical properties of silicon carbide and for his valuable support 
with the simulation codes. This work was funded by the Ministerio de 
Ciencia, Innovaci\'on y Universides (Spain) under Grant 
PID2022-139776NB-C66 and by 
the ``Severo Ochoa Centres of Excellence'' program under Grant 
CEX2024-001445-S. We also acknowledge the CINECA award within the ISCRA 
initiative (Italy) for providing high-performance computing resources 
and technical support.  \\  \\
\end{acknowledgments}

\noindent
{\bf Data availability} \\

The data that support the findings of this article are openly
available \cite{sc-ze26}.  \\ \\

%  -------------------------------------------------------------------

%  BIBLIOGRAPHY

%

\end{document}